\pdfoutput=1
\documentclass[3p,preprint,11pt]{elsarticle}

\usepackage[T1]{fontenc}
\usepackage[utf8]{inputenc}
\usepackage{amsmath}
\usepackage{amssymb}
\usepackage[title]{appendix}

\usepackage[all]{nowidow}

\usepackage{booktabs}
\usepackage{makecell}

\usepackage{subfig}

\usepackage[makeroom]{cancel}

\usepackage{siunitx}
\usepackage[ruled,vlined,linesnumbered]{algorithm2e}
\SetKwComment{Comment}{$\triangleright$\ }{}

\usepackage{amsthm}
\theoremstyle{remark}
\newtheorem{remark}{Remark}

\usepackage{float}
\usepackage{placeins}

\usepackage{changes}
\makeatletter
\@namedef{Changes@AuthorColor}{red}
\colorlet{Changes@Color}{red}
\makeatother

\usepackage{./commands}
\usepackage{./colors}
\usepackage{./symbols}

\journal{Elsevier}

\newcommand{\theTitle}{
A hybrid sharp--diffuse interface approach to accurately model melt pool dynamics with rapid evaporation in laser-based processing of metals
}

\usepackage{hyperref}
\hypersetup{
	pdfinfo={
		Title={\theTitle},
		Author={Nils Much, Andreas Koch, Christoph Meier, Magdalena Schreter-Fleischhacker}
	},
	colorlinks
}

\biboptions{numbers,sort&compress}

\begin{document}

\begin{frontmatter}

\title{\theTitle}

\author[SAM]{Nils Much\corref{cor1}}
\cortext[cor1]{Corresponding author}

\author[SAM]{Andreas Koch}
\author[SAM]{Christoph Meier}
\author[SAM]{Magdalena Schreter-Fleischhacker}

\affiliation[SAM]{
	organization={Professorship~of~Simulation~for~Additive~Manufacturing,~Technical~University~of~Munich},
 	addressline={\! Freisinger~Landstraße~52},
 	city={Garching~b.~München},
 	postcode={85748},
 	country={Germany}
}

\begin{abstract}

Predictive simulation of melt pool dynamics in laser-based processing of metals, e.g., laser beam welding or laser powder bed fusion additive manufacturing, requires accurate resolution of thermo-hydrodynamic interactions at the melt--gas interface.
Here, evaporation-induced recoil pressure and temperature-dependent surface tension govern the flow.
Because these mechanisms depend sensitively, often exponentially, on the interface temperature, reliable predictions demand highly accurate heat transfer models.
Popular diffuse-interface formulations smear the extreme thermal gradients as typical for laser--metal interactions, leading to interface temperature errors that critically degrade the accuracy of interface force predictions and melt pool dynamics.
We present a hybrid sharp--diffuse interface approach for high-fidelity modelling of melt pool thermo-hydrodynamics with rapid evaporation.
The heat transfer problem is represented using a sharp-interface unfitted finite element (CutFEM) formulation, enabling accurate prediction of the temperature field.
The multi-phase flow problem, characterized by large density ratios and complex interface dynamics, is accurately captured using a robust level-set-based one-fluid diffuse-interface finite element formulation.
Consistent coupling is achieved by extending the sharp-interface temperature into a narrow interface region to evaluate temperature-dependent interface forces within the diffuse-interface flow framework.
In practically relevant benchmarks, the sharp-interface thermal model exhibits second-order spatial convergence, enabling finite element sizes two orders of magnitude larger than the diffuse-interface approach for 1\% accuracy.
In a novel coupled thermo-hydrodynamic benchmark representative of laser--metal interactions, the hybrid approach is one order of magnitude more accurate than a purely diffuse-interface model on the same mesh.
Robustness is demonstrated in a 3D proof-of-principle simulation of stationary laser-induced melting of a bare metal plate.
The methodology applies to basically all problems with temperature-driven flow, in particular when strongly localized heat sources are involved, as, e.g., in laser beam welding and laser powder bed fusion.

\end{abstract}


\end{frontmatter}

\section{Introduction}
\label{sec:intro}

\subsection{Background and challenges}

Laser-based processing of metals, such as laser beam welding and additive manufacturing via laser powder bed fusion (\PBFAM), offers promising capabilities for producing high-performance parts~\cite{gibson2021additive}.
These processes are governed by a complex interplay of physical mechanisms spanning multiple length and time scales~\cite{meier2017thermophysical}.
In the vicinity of the laser, melt pool dynamics are governed by a complex, highly dynamic multi-phase thermo-hydrodynamic problem with a priori unknown, moving interfaces, complex topology changes, and large parameter jumps ($\sim\num{e2}$ to $\sim\num{e5}$) at the melt--gas boundary~\cite{li2022particle}.
The strongly localized laser superheats the melt far beyond its (equilibrium) boiling temperature, leading to extreme temperature gradients and rapid phase change, including melting, solidification, and non-equilibrium evaporation.
The resulting evaporation-induced recoil pressure emerges as a major driving force for melt pool dynamics and can create deep vapor depressions, so-called keyholes, under typical process conditions~\cite{cunningham2019keyhole}.
Depending on the process conditions, strongly fluctuating keyholes at the melt pool front may foster the entrainment of gas bubbles that can solidify as pores in the final part, a phenomenon denoted as keyhole instability.
Thereby, the intensity of evaporation effects is highly sensitive to the melt pool surface temperature, necessitating highly accurate temperature prediction.
In contrast, flow at the tail end of the melt pool is often surface-tension dominated: long, slender melt pool shapes combined with poor wettability promote Rayleigh--Plateau instabilities, i.e., the continuous melt track breaks into single droplets (balling), which significantly deteriorates the surface quality of the final part; accordingly, temperature-dependent surface tension strongly influences the melt pool dynamics~\cite{korner2013fundamental} and is crucial for predicting undesired balling.

Process-induced porosity and related defects are closely linked to melt pool instabilities such as keyholing, lack of fusion, balling, and denudation~\cite{brennan2021defects}.
In practice, their mitigation still necessitates extensive parameter tuning via trial and error, reflecting the absence of a predictive, unified computational framework that links defect formation to process parameters and material properties.
While computational modelling has emerged as a crucial tool, offering simulation-based insights into the complex physics of laser-based metal processing, a computational melt pool model that can accurately resolve the complex, temperature-sensitive process interactions across practically relevant time and length scales remains pending.
Developing such a high-fidelity computational melt pool model is urgently needed to navigate and exploit new processing regimes that enable higher throughput without compromising quality, and is therefore the central objective of this work.

\subsection{Related work on computational melt pool models}

A predictive simulation tool for melt pool dynamics must combine physically sound mathematical multi-phase models with numerical schemes that are simultaneously accurate, robust, and computationally efficient.
Although a wide range of melt pool simulation approaches has been proposed in recent years~\cite{cook2020simulation}, existing studies consistently show that quantitatively reliable predictions either require prohibitively high computational costs or rely on extensive (artificial) calibration of poorly constrained parameters.
A central modelling challenge arises from the transport of mass, momentum, and energy across the rapidly evolving metal--gas interface.
Particularly, accurate prediction of evaporation-induced recoil pressure and cooling, as well as temperature-dependent surface tension, which are the defining driving forces of melt pool dynamics, especially in the keyhole regime, critically depends on precise resolution of the interface temperature and its gradients.
Consequently, the representation of the metal--gas interface emerges as a key modelling choice that strongly influences both predictive accuracy and computational feasibility.
Existing melt pool models can be broadly categorized into sharp-interface (SI) and diffuse-interface (DI) approaches.

SI approaches maintain discontinuities and are therefore well-suited for high-fidelity modelling of interface problems.
For Eulerian, unfitted frameworks, popular methods include the \underline{Cut} \underline{F}inite \underline{E}lement \underline{M}ethod (CutFEM)~\cite{hansbo2002unfitted}, XFEM~\cite{chessa2002extended}, and the ghost-fluid method~\cite{fedkiw1999nonoscillatory}, where the accuracy gains require modifications of the numerical schemes, such as stabilization for the small cut-element problem~\cite{burman2010ghost} or weak imposition of coupling conditions for interfaces not aligned with the mesh.
In alternative fitted mesh methods, such as arbitrary Lagrangian--Eulerian (ALE)~\cite{hirt1974arbitrary} approaches or particle-FEM~\cite{fevrier2025simulation}, frequent remeshing is necessary.
While this is manageable if interface movement is known a priori, melt pool dynamics typically involve highly dynamic, complex interface topology changes.
Unfitted approaches require reconstruction of the interface and intersected quadrature generation for every interface movement, but this computational overhead remains comparatively constant even for complex interface movements.
Therefore, unfitted SI approaches, such as CutFEM, appear better suited to melt pool dynamics compared to ALE approaches, which encounter difficulties in such scenarios.
Furthermore, these SI multi-phase flow models often require special methods to reliably represent breakup and coalescence of the interface topology~\cite{tryggvason2011direct}.
While these methods offer high physical fidelity, their numerical complexity and limited robustness have restricted SI melt pool simulations to only a few formulations, notably finite-difference (FDM) / ghost-fluid approaches \mbox{\cite{pang2011three, tan2013investigation}}, and hybrid finite element (FEM) -- finite volume (FVM) frameworks employing an ALE description \mbox{\cite{khairallah2014mesoscopic, khairallah2016laser, matthews2016denudation, ly2017metal, martin2019dynamics}}.

Alternatively, allowing for a more straightforward and robust implementation, DI methods regularize discontinuities across a finite thickness, often using the Continuum Surface Force (CSF) model~\cite{brackbill1992continuum} for surface tension and extensions for heat fluxes~\cite{much2024improved}.
While less accurate compared to SI approaches, DI approaches are mathematically consistent, converging to sharp-interface solutions as the thickness decreases.
They also robustly handle topology changes and integrate with standard single-phase solvers, as long as these solvers support variable material parameters and the augmentation of right-hand-side terms, which explains their frequent use in modelling melt pool dynamics.
DI melt pool models include formulations based on FEM \mbox{\cite{much2024improved, schreter2025consistent, andreotta2017finite, yan2018fully, zhu2021mixed, courtois2014complete, leitz2018fundamental, queva2020numerical}}, FVM \mbox{\cite{panwisawas2017mesoscale, bayat2019multiphysics, grohol2024predictive, yu2024mechanism, geiger20093d, chen2020spattering}}, FDM~\cite{lee2015mesoscopic}, lattice Boltzmann \mbox{\cite{korner2013fundamental, zakirov2024kissam, ikeda2025high, ammer2014simulating}}, and smooth particle hydrodynamics \mbox{\cite{meier2021novel, fuchs2022versatile, luthi2023adaptive, lin2023enhanced, ma2025gpu, russell2018numerical}}.

Furthermore, melt pool models are distinguished by their type of evaporation model:
\emph{Resolved} evaporation models consider the exchange of mass from liquid metal to metal vapor, e.g.~\cite{schreter2025consistent}, while the standard approach is the usage of \emph{unresolved} evaporation models that neglect mass exchange and vapor flow and instead consider an approximate analytical model for the evaporation-induced pressure jump~\cite{anisimov1995instabilities} and cooling at the melt--gas interface.
Nevertheless, both approaches require a highly accurate interface temperature prediction.

The predictive fidelity of computational melt pool models critically depends on accurately resolving the thermal field and its gradients at the metal--gas interface, as evaporation-induced recoil pressure and cooling, which are the primary driving forces for melt pool dynamics, scale exponentially with the local melt pool surface temperature.
Thus, small (discretization) errors in the temperature field can lead to drastic errors in the predicted melt pool dynamics.
Despite their frequent use, DI methods smear interface jumps, such as those of temperature gradients, introducing critical modelling and discretization errors in the temperature field~\cite{much2024improved}.
The resulting deviation from experimental measurements, as discussed in detail, e.g., in \mbox{\cite{khairallah2016laser, andreotta2017finite, zhu2021mixed, ross2022volumetric}}, is often addressed by model calibration, i.e., fitting model parameters such as the laser absorptivity, such that simulation results better align with experimental measurements.
However, this simple approach cannot compensate for the underlying problem, i.e., an insufficient interface resolution, in general.
Therefore, our recent work~\cite{much2024improved} showed that SI thermal models are urgently needed for high-fidelity predictions of melt pool dynamics, as DI models would require extremely fine mesh resolutions, thus resulting in impermissibly high computational costs to achieve an acceptable level of accuracy.

\subsection{Our contributions}

Virtually no research addresses the combined use of sharp and DI methods for complex multi-phase, multi-physics problems such as melt pool dynamics.
Such a hybrid approach is particularly promising in this setting, where thermal fields demand high precision at the interface, while flow fields -- especially with extreme interface forces and complex interface dynamics -- are more robustly handled using DI methods.
To close this gap, we present the first hybrid sharp--diffuse interface (HSDI) approach for two-phase thermo-hydrodynamics that couples SI CutFEM temperature modelling with DI FEM flow modelling, based on an unresolved evaporation model, achieving an unprecedented balance of predictive accuracy and computational efficiency.
Departing from a well-established DI two-phase flow model based on a mass-conserving level-set framework~\cite{olsson2007conservative} used in our previous works on melt pool dynamics \mbox{\cite{much2024improved, schreter2025consistent}}, we outline the main contributions as two-fold:

First, we develop a CutFEM-based SI formulation that enables highly accurate predictions of the temperature field, particularly at the melt pool surface.
We investigate two variants tailored to melt pool dynamics.
The first is a coupled two-phase formulation defined over the full metal--gas domain, in which temperature continuity across the metal--gas interface is weakly enforced using Nitsche’s method~\cite{nitsche1971uber}.
The second is a reduced and thus more efficient single-phase formulation restricted to the metal domain, motivated by the negligible heat capacity of the ambient gas.
In this case, the metal--gas interface is treated as adiabatic except for the heat fluxes induced by the laser heat source and evaporative cooling.

Second, we propose a consistent coupling strategy between the SI thermal model and the DI flow model.
Since key interface forces (recoil pressure, surface tension) depend sensitively on temperature, we extend the SI temperature field into a narrow band surrounding the metal--gas interface using a closest-point projection.
This enables accurate and consistent evaluation of temperature-dependent interface source terms within the DI framework.

Furthermore, we propose a new benchmark example, resembling a \emph{laser-induced vapor depression} in 2D, tailored to investigate the accuracy of the proposed and existing melt pool models in comprehensive convergence studies with practically relevant error measures and in a setting representative of melt pool dynamics.
In this benchmark example, the proposed HSDI approach turns out to be one order of magnitude more accurate than a standard DI model on the same mesh.

With the HSDI approach, we aim to provide an efficient computational model that enables unprecedented accuracy for melt pool simulations on mesh resolutions feasible for industrially relevant length and time scales.
To this end, we leverage state-of-the-art high-performance computing techniques, including matrix-free operator evaluation~\cite{kronbichler2012generic}, adaptive mesh refinement, and MPI-based parallelization using domain decomposition provided by the \texttt{deal.II} library~\cite{arndt2025deal}, which is suitable for exascale computing.

The remainder of this article is structured as follows:
Section~\ref{sec:methods} presents the governing equations of the underlying thermo-hydrodynamics model with phase change, the HSDI description including the two SI heat transfer model variants and their coupling with the DI two-phase flow model using the extension of the interface temperature, and the numerical framework.
In Section~\ref{sec:numerical_examples}, the accuracy of the presented approaches is verified using selected benchmark examples and compared to competing melt pool models.
Furthermore, we demonstrate the capabilities of the presented methods using a practically relevant example of stationary laser illumination.
Conclusions are drawn in Section~\ref{sec:conclusion}.

\section{Methods}
\label{sec:methods}

\begin{figure}[b!]
	\centering
	\includegraphics{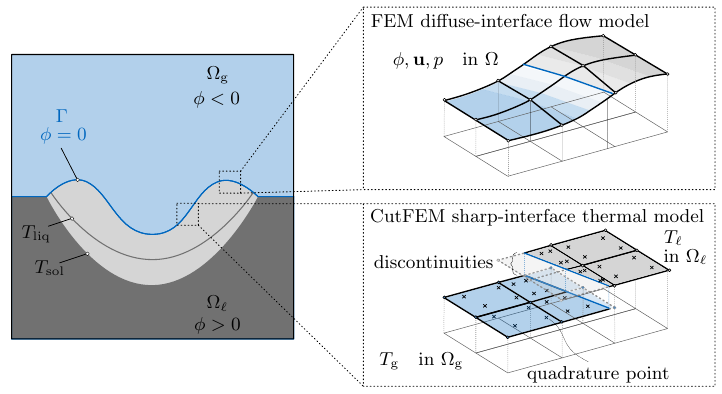}
	\caption{
		Qualitative sketch of the domain and the computational HSDI thermal two-phase flow model for melt pool modelling.
		(left)~The physical domain of interest~$\Omega$ comprises a metal phase~$\OmegaL$ and an ambient gas phase~$\OmegaG$.
		The metal phase is further distinguished into a melt and a solid phase, which are separated by the mushy zone between the liquidus temperature~$\Tliquidus$ and the solidus temperature~$\Tsolidus$.
		(top~right)~For the two-phase flow submodel, a FEM-based DI approach is used, where properties and interface forces between phases are smeared smoothly in a diffuse transition region centred around the interface~$\Gamma$.
		(bottom~right)~For the two-phase heat transfer submodel, a CutFEM-based SI approach is used, where discontinuities in the temperature field are accurately resolved.
	}
	\label{fig:hybrid_melt_pool_sketch}
\end{figure}

Motivated by the accuracy limitations of purely DI melt pool dynamics models, as discussed in our previous work~\cite{much2024improved}, we present a new HSDI approach for thermal two-phase flow.
This model is specifically tailored to the stringent requirements of laser-induced melt pool modelling with rapid evaporation:
While the thermal field demands the highest precision at the melt pool surface, as provided by a SI CutFEM approach, the complex interfacial flow dynamics can be captured robustly by the DI fluid formulation.
Our mathematical model builds upon the following assumptions:
\begin{itemize}
	\item Convective and conductive heat transfer, neglecting contributions due to viscous or capillary dissipation.
	\item Jump in thermal properties and a sharp representation of thermal fluxes at the interface.
	\item Immiscible, incompressible, viscous (Newtonian) two-phase flow at moderate Reynolds numbers.
	\item Diffuse transition region at the interface for flow properties and interface forces between the phases, centred around the (sharp) interface with a finite but small thickness.
	\item Rigid and immobile solid phase.
\end{itemize}
In the left panel of Fig.~\ref{fig:hybrid_melt_pool_sketch}, the domain of interest and the composition of the phases are shown.
We consider an Eulerian domain ${\Omega = \OmegaL \cup \OmegaG \subset \{\Bx \in \Real^{d}\}}$ with dimensionality ${d \in \{1,2,3\}}$, which is divided into a metal phase~$\OmegaL$ and an ambient gas phase~$\OmegaG$ with ${\OmegaL \cap \OmegaG = \emptyset}$.
The metal phase~$\OmegaL$ comprises a melt and a solid phase with reversible phase change of melting and solidification, as well as a diffuse transition region in the mushy zone between the liquidus temperature~$\Tliquidus$ and solidus temperature~$\Tsolidus$ isosurfaces.
The metal--gas interface ${\Gamma = \partial\OmegaL \cap \partial\OmegaG \in \Real^{d-1}}$ is exposed to laser-induced heating, evaporation-induced recoil pressure and cooling, and temperature-dependent surface tension.

\subsection{Governing equations}
\label{sec:governing_eq}

The following presents the governing equations that comprise the underlying mathematical model for melt pool simulations.
The thermal two-phase flow is governed by the incompressible Navier--Stokes equation and the heat equation, along with their respective coupling conditions.
Additionally, a level-set method is used for interface tracking.

\subsubsection{Level-set}
\label{sec:governing_level_set}

The metal--gas interface~$\Gamma$ is captured as the zero-isosurface of a regularized level-set function ${\levelset := \levelset(\Bx, t)}$ with ${\levelset \in [-1, 1]}$ according to \mbox{\cite{olsson2007conservative, kronbichler2018fast}}.
We denote ${\levelset > 0}$ as inside the metal phase~$\OmegaL$ and ${\levelset < 0}$ as inside the ambient gas phase~$\OmegaG$.
The temporal evolution of the level-set function~$\levelset$ is governed by the advection equation
\begin{align} \label{eq:advection_equation}
	\fracPartial{\levelset}{t} + \vel\cdot\nabla\levelset = 0
	\quad\text{ in }\Omega\times[0,t]
\end{align}
with the fluid velocity field ${\vel := \vel(\Bx,t)}$ and time~$t$.
The initial condition of the level-set function is determined from the signed distance ${\distance := \distance(\Bx, t)}$ to the initial metal--gas interface according to
\begin{align} \label{eq:initial_levelset}
	\levelset(\Bx) = \tanh\left( \frac{3\,\distance(\Bx)}{\interfaceThickness} \right),
\end{align}
where~$\interfaceThickness$ is the thickness of the diffuse transition region, which we refer to as the interface thickness.
To maintain the regularized characteristic shape of the level-set function, we perform an algebraic reinitialization procedure after every advection time step, as described in~\cite {olsson2007conservative}.

For the interpolation of physical properties between the phases, we define a localized, smoothed indicator function ${\indicator := \indicator(\Bx, t)}$ with ${\indicator \in [0, 1]}$ for which we choose a sine-based smoothed Heaviside function according to \mbox{\cite{sussman1994level, peskin2002immersed}}:
\begin{align}
	\indicator(\distance) &=
	\begin{cases}
		0 & \text{for} \quad \distance \leq -\frac{\interfaceThickness}{2} \\
		\frac{1}{2} + \frac{\distance}{\interfaceThickness} + \frac{1}{2 \pi} \sin{\left(\frac{2 \pi \distance}{\interfaceThickness}\right)} & \text{for} \quad -\frac{\interfaceThickness}{2} < \distance < \frac{\interfaceThickness}{2} \\
		1 & \text{for} \quad \distance \geq \frac{\interfaceThickness}{2}
	\end{cases}
	\label{eq:indicator} \\
	\text{with} \quad
	\distance(\levelset) &= \frac{\interfaceThickness}{6} \log\left( \frac{1 + \levelset}{1 - \levelset} \right)
	\nonumber
\end{align}
\pagebreak

\subsubsection{Heat transfer}
\label{sec:governing_heat}

In the metal phase~$\OmegaL$ and ambient gas phase~$\OmegaG$, the evolution of the temperature ${T := T(\Bx,t)}$ is governed by the heat equation, including conduction and convection
\begin{align}\label{eq:heat_equation_two_phase}
	\rhoS \cpS \left( \frac{\partial \Ts}{\partial t} + \vel \cdot \nabla \Ts \right)
	= \nabla \left( \condS \, \nabla \Ts \right)
	\quad \text{in }\OmegaS \times [0,t]
	\quad \text{for } \phaseIndex \in \metalAndGas
	.
\end{align}
The density~$\rho$, specific heat capacity~$\cp$ and thermal conductivity~$\cond$ are phase-specific thermal properties which have the subscript ${(\bullet)_\phaseIndex := \inLiquid{(\bullet)}}$ for the metal phase~$\OmegaL$ and the subscript ${(\bullet)_\phaseIndex := \inGas{(\bullet)}}$ for the ambient gas phase~$\OmegaG$.
The heat equation~\eqref{eq:heat_equation_two_phase} is supplemented by an initial condition
\begin{alignat}{3}
	\Ts &= T_0
	&&\quad \text{in } \OmegaS \times \{t=0\}
	&&\quad \text{for } \phaseIndex \in \metalAndGas
	\label{eq:inital_condition_two_phase}, \\
\intertext{
as well as Dirichlet and Neumann boundary conditions on the outer boundary ${\partial\Omega = \GammaD \cup \GammaN}$ with ${\GammaD \cap \GammaN = \emptyset}$
}
	\Ts &= \hat{T}
	&&\quad \text{on } \GammaD \cap \partial\OmegaS \times [0,t]
	&&\quad \text{for } \phaseIndex \in \metalAndGas
	\label{eq:Dirichlet_two_phase} \\
	-\condS \, \partial_{\outerNormal} \Ts &= \qHat
	&&\quad \text{on } \GammaN \cap \partial\OmegaS \times [0,t]
	&&\quad \text{for } \phaseIndex \in \metalAndGas
	\label{eq:outer_Neumann_two_phase},
\end{alignat}
where $\hat{T}$ denotes the prescribed temperature at the boundary, $\qHat$ denotes the prescribed heat flux at the boundary, and~$\partial_{\outerNormal}$ is the directional derivative in the normal direction, indicated by the outward-pointing unit normal vector~$\outerNormal$ at the domain boundary~$\partial\Omega$.
At the metal--gas interface~$\Gamma$, a Dirichlet-type interface matching condition and a heat flux jump condition are imposed:
\begin{alignat}{2}
	\left[ T \right]_\Gamma
	&= 0
	&&\quad \text{on } \Gamma \times [0,t]
	\label{eq:interface_matching_two_phase} \\
	\left[ -\cond \, \partial_{\interfaceNormal} T \right]_\Gamma
	&= \qGamma
	&&\quad \text{on } \Gamma \times [0,t]
	\label{eq:interface_matching_heat_flux_two_phase}
\end{alignat}
Here, $\left[ \bullet \right]_\Gamma$ denotes the interface jump operator ${\left[ x \right]_\Gamma = \left.\inLiquid{x}\right|_\Gamma - \left.\inGas{x}\right|_\Gamma}$, and~$\partial_{\interfaceNormal}$ is the directional derivative in the normal direction of the interface.
The interface unit normal vector~$\interfaceNormal$ points into~$\OmegaL$.
We split the interface heat flux according to
\begin{align} \label{eq:laser_plus_evapor_heat_loss}
	\qGamma = \qLaser + \qVapor(\TGamma)
	\text{,}
\end{align}
which comprises the laser heat source~$\qLaser$, the detailed form of which is defined in Section~\ref{sec:numerical_examples} for specific examples, and the temperature-dependent evaporation-induced cooling~$\qVapor$
\begin{align} \label{eq:evaporative_heat_loss_with_specific_enthalpy}
	\qVapor(\TGamma) = -h(\TGamma)\,\mDot(\TGamma)
	\quad \text{ for } \TGamma \geq \Tv
	\quad \text{with} \quad
	h(\TGamma) = \hv + \int_{\Thvref}^{\TGamma}\cp\,\diffd T,
\end{align}
where $h(\TGamma)$ is the specific enthalpy of the vapor, and $\Tv$ is the boiling temperature at the ambient pressure $\pa$.
The interface temperature~$\TGamma$ is defined by the continuity condition~\eqref{eq:interface_matching_two_phase} as ${\TGamma = \left.\Tl\right|_\Gamma = \left.\Tg\right|_\Gamma}$, where $\left.\Tl\right|_\Gamma$ and $\left.\Tg\right|_\Gamma$ indicate the temperature of the metal and ambient gas phase at the interface.
In this work, we consider an unresolved evaporation model and thus neglect the explicit exchange of mass from liquid to vapor.
Instead, the effects due to evaporative phase change at the melt--vapor interface are accounted for by phenomenological formulations for thermal and mechanical fluxes, which is the standard approach for melt pool models \mbox{\cite{khairallah2016laser, leitz2018fundamental, queva2020numerical, zhu2021mixed, courtois2014complete, panwisawas2017mesoscale, chen2020spattering, lee2015mesoscopic, zakirov2024kissam, ikeda2025high, meier2021novel, fuchs2022versatile, luthi2023adaptive, lin2023enhanced, ma2025gpu}}.
As a consequence, convective heat transfer in the gas phase resulting from the mass exchange across the interface is neglected.
The expression for the evaporation-induced cooling~$\qVapor$ is adapted to account for the enthalpy transported by the phenomenologically modelled vapor mass flux, considering the specific enthalpy of the vapor~$h(\TGamma)$, which can be split into the latent heat of evaporation~$\hv$ and a purely temperature-dependent contribution, which can be expressed via the specific heat capacity $\cp$ and its reference temperature~$\Thvref$, according to \eqref{eq:evaporative_heat_loss_with_specific_enthalpy}.

We determine the phenomenological vapor mass flux~$\mDot$ at the metal--gas interface~$\Gamma$ using the model proposed by Knight~\cite{knight1979theoretical} according to
\begin{align} \label{eq:evaporative_mass_flux}
	\mDot(\TGamma) = 0.82\,\cs\,\pv\,\sqrt{\frac{M}{2\pi\,R\,\TGamma}}
	\quad \text{ for } \TGamma \geq \Tv
\end{align}
with the molar mass~$M$ and the molar gas constant~$R$.
For metals, the sticking constant is typically assumed as ${\cs = 1}$~\cite{khairallah2016laser}.
In Knight's model~\cite{knight1979theoretical}, the vapor mass flux is computed based on the saturated vapor pressure, while it is directly formulated as a function of the recoil pressure~$\pv$ in our simplified model.
The evaporation-induced recoil pressure~$\pv$ is determined by the phenomenological model by Anisimov and Khokhlov~\cite{anisimov1995instabilities}:
\begin{align} \label{eq:recoil_pressure}
	\pv(\TGamma) = 0.54\,\pa\,\exp\left(-\frac{\hvBar}{R}\left(\frac{1}{\TGamma}-\frac{1}{\Tv}\right)\right)
	\quad \text{ for } \TGamma \geq \Tv
\end{align}
Here, $\hvBar$ is the molar latent heat of evaporation, and ${\pa = \SI{e5}{Pa}}$ is the atmospheric pressure.

\begin{figure}[hbt!]
	\centering
	\includegraphics{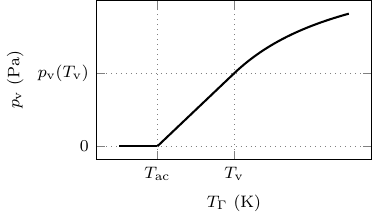}
	\caption{
		Recoil pressure~$\pv(\TGamma)$~\eqref{eq:recoil_pressure} over time, considering the linear activation function for ${\TGamma\leq \Tv}$ according to~\eqref{eq:ramp_up}.
	}
	\label{fig:recoil_pressure_ramp}
\end{figure}

\begin{remark}
With the formulations in~\eqref{eq:evaporative_heat_loss_with_specific_enthalpy},~\eqref{eq:evaporative_mass_flux}, and~\eqref{eq:recoil_pressure}, the evaporation-induced cooling~$\qVapor$, mass flux~$\mDot$, and recoil pressure~$\pv$ exhibit a jump at the boiling temperature~$\Tv$, which results in difficulties for the numerical solution procedure.
To mitigate this issue, we introduce an activation temperature ${\Tac < \Tv}$, from which we ramp up those quantities (${f\in\{\qVapor, \mDot, \pv\}}$) between $\Tac$ and $\Tv$ via
\begin{align} \label{eq:ramp_up}
	f(\TGamma) = \max\left( f(\Tv) \frac{\TGamma - \Tac}{\Tv - \Tac} , 0 \right)
	\quad \text{ for } \TGamma < \Tv,
\end{align}
such that ${f(\TGamma)=0}$ holds for ${\TGamma \le \Tac}$ and $f(\TGamma)$ increases linearly to $f(\Tv)$ for ${\Tac\leq \TGamma \leq\Tv}$.
As an example, in Fig.~\ref{fig:recoil_pressure_ramp}, the resulting recoil pressure temperature dependency is shown.
\end{remark}
\pagebreak

\subsubsection{Two-phase flow}
\label{sec:governing_flow}

The two-phase flow is modelled using a one-fluid DI approach, as shown in the top right panel of Fig.~\ref{fig:hybrid_melt_pool_sketch}.
In a one-fluid approach, the two material phases are treated as a single fluid with variable properties, which are determined based on the indicator function~$\indicator(\Bx)$~\eqref{eq:indicator}, which is provided by the level-set framework.
The flow field is governed by the incompressible Navier--Stokes equations, composed of the momentum balance equation and continuity equation
\begin{alignat}{2}
	\rhoEff \left(\fracPartial{\vel}{t} + \left( \vel\cdot\nabla\right) \vel \right) &= -\nabla p + \nabla\cdot\left( 2\muEff\,\Bvep \right) + \fRecoilPressureDiffuse + \fSurfaceTensionDiffuse + \fDarcy
	&&\quad \text{in }\Omega\times[0,t]
	\label{eq:momentum_equation} \\
	\nabla\cdot\vel &= 0
	&&\quad \text{in }\Omega\times[0,t]
	\label{eq:continuity_equation}
\end{alignat}
with the pressure field ${p := p(\Bx,t)}$, effective density~$\rhoEff$, effective dynamic viscosity~$\muEff$, rate-of-deformation tensor $\Bvep = \frac{1}{2}(\nabla\vel + (\nabla\vel)^\top)$, volume forces resulting from the evaporation-induced recoil pressure force~$\fRecoilPressureDiffuse$ and surface tension~$\fSurfaceTensionDiffuse$, as well as the Darcy damping force~$\fDarcy$.
The equations~\eqref{eq:momentum_equation} and~\eqref{eq:continuity_equation} are supplemented by an initial condition, slip, and no-slip boundary conditions
\begin{alignat}{2}
	\vel &= \vel_0
	&&\quad \text{in } \Omega \times \{t=0\}
	\label{eq:inital_condition_velocity} \\
	\vel \cdot \outerNormal &= 0
	&&\quad \text{on } \GammaSlip \times [0,t]
	\label{eq:slip_BC} \\
	\vel &= \Bzero
	&&\quad \text{on } \GammaNoSlip \times [0,t]
	\label{eq:no_slip_BC}
\end{alignat}
with ${\GammaSlip \cap \GammaNoSlip = \partial\Omega}$ and ${\GammaSlip \cup \GammaNoSlip = \emptyset}$.
In addition, a zero-pressure boundary condition is applied at a single boundary point to normalize the pressure field.
The effective physical flow properties, denoted by the subscript~$\interp{(\bullet)}$, are assumed to be constant within the bulk region of the individual (liquid/solid/gas) phases.
Only in the diffuse-interface transition region, these effective fluid properties ${\eff{\alpha}\in\{\rho, \mu\}}$ are computed from the arithmetic mean of their phase contributions
\begin{align} \label{eq:parameter_transition_arithmetic}
	\eff{\alpha}(\indicator) = \inLiquid{\alpha}\,\indicator + \inGas{\alpha}(1-\indicator),
\end{align}
considering the indicator function $\indicator$~\eqref{eq:indicator}.

We formulate our model for melting and solidification based on the simplifying assumption that the solid phase is rigid and immobile.
To that, we artificially increase the viscosity and employ a Darcy damping force~\cite{voller1987fixed} in the solid phase to inhibit its motion, adopted from \mbox{\cite{much2024improved, schreter2025consistent}}.
Note that comparative studies suggest that only employing a Darcy damping force would be sufficient here, resulting in the same material behaviour for the limit of high damping coefficients.
For the dynamic viscosity in the metal phase~$\inLiquid{\mu}$, we interpolate between the molten and solid metal phase according to
\begin{align} \label{eq:metal_viscosity}
	\inLiquid{\mu}(\solidFraction) = \inMelt{\mu} + \left( \inSolid{\mu} - \inMelt{\mu} \right) \left( -2 \solidFraction^3 + 3\solidFraction^2 \right),
\end{align}
using the temperature-dependent solid fraction
\begin{align} \label{eq:solid_fraction}
	\solidFraction(T,\indicator) = \begin{cases}
		\indicator &\text{for} \quad T \leq \Tsolidus \\
		\indicator\,\frac{\Tliquidus - T}{\Tliquidus - \Tsolidus} &\text{for} \quad \Tsolidus < T < \Tliquidus \\
		0 &\text{otherwise}
	\end{cases}.
\end{align}
The temperature range between the liquidus temperature~$\Tliquidus$ and the solidus temperature~$\Tsolidus$ defines the mushy zone, where the solid fraction transitions from \num{0} to \num{1}.
The remaining properties (density, thermal conductivity, and specific heat capacity) are assumed to be the same for the molten and solid metal phase.
Eq.~\eqref{eq:metal_viscosity} enables us to assign a large viscosity to the solid phase, which suppresses spurious velocity gradients and, in combination with the Darcy damping force, described below, effectively represents our melting and solidification model, considering a rigid, immobile solid metal.
The Darcy damping force~$\fDarcy$ in~\eqref{eq:momentum_equation} is determined according to
\begin{align} \label{eq:darcy_damping_term}
	\fDarcy(\vel,\solidFraction) = -\darcyMorthology\left( \frac{\solidFraction^2}{(1 - \solidFraction)^{3} + \darcyDivZero} \right) \vel
\end{align}
with the parameter~$\darcyMorthology$, which depends on the mushy zone morphology, and a parameter~$\darcyDivZero$ to avoid division by zero~\cite{voller1987fixed}.

In~\eqref{eq:momentum_equation}, regularized interface forces are marked by a superscript tilde~$(\tilde{\bullet})$, which include the evaporation-induced recoil pressure force~$\fRecoilPressureDiffuse$ and the surface tension force~$\fSurfaceTensionDiffuse$.
The corresponding equations will be specified in Section~\ref{sec:diffuse_two_phase_flow_model}.

\subsection{Discrete weak forms of the governing equations}
\label{sec:weak_form}

The governing equations comprise the heat equation~\eqref{eq:heat_equation_two_phase}, the incompressible Navier--Stokes equations~\eqref{eq:momentum_equation} and~\eqref{eq:continuity_equation}, the advection equation~\eqref{eq:advection_equation}, and additional equations for the level-set framework, which include the reinitialization~\cite{olsson2007conservative}, as well as regularization of normal vector and curvature.
A detailed description of the additional level-set equations is provided in~\cite{schreter2024consistent}.

For the solution of the governing partial differential equations, we employ one Eulerian mesh with hexahedral continuous finite elements.
The finite element side length~$h$ is defined as ${h = \sqrt[d]{h_\*{max}}}$, where~$h_\*{max}$ is the longest element diagonal and~$d$ is the spatial dimension.
We utilize a Bubnov--Galerkin ansatz and Lagrange polynomial shape functions for spatial discretization.
Specifically, we use quadratic shape functions for the velocity field and linear shape functions for the pressure field to ensure inf-sup stability.
We also use linear shape functions for the level-set framework and for the heat transfer solver.
However, compared to the heat transfer and flow problem, we support refining the mesh of the level-set framework by subdividing each finite element once in each spatial dimension.
This approach helps to improve accuracy, and in combination with the level-set reinitialization procedure~\cite{olsson2007conservative}, spurious oscillations of the level-set and interface topology are minimized.
Thus, at moderate Reynolds numbers, no additional stabilization of the level-set advection is required~\cite{kronbichler2018fast}.

\subsubsection{Navier--Stokes equations}
\label{sec:weak_flow}

The strong form of the Navier--Stokes equations~\eqref{eq:momentum_equation} and~\eqref{eq:continuity_equation} is transferred into the weak form by the method of weighted residuals.
\sloppy{
The solution space for the velocity is defined as ${\mathcal{S}_\vel = \{ \vel \in \left( H^1(\Omega) \right)^d\, |\, \vel\cdot\outerNormal = 0 \text{ on } \GammaSlip\, |\, \vel = \Bzero \text{ on } \GammaNoSlip\}}$ and for the corresponding test functions as ${\mathcal{V}_\vel = \{ \vu \in \left( H^1(\Omega) \right)^d\, |\, \vu\cdot\outerNormal = 0 \text{ on } \GammaSlip \,|\, \vu = \Bzero \text{ on } \GammaNoSlip\}}$, considering the slip~\eqref{eq:slip_BC} and no-slip~\eqref{eq:no_slip_BC} boundary conditions.
}
Here, ${H^1(\Omega)}$ represents the space of square integrable functions with square integrable derivatives.
The solution spaces for the pressure and its test functions are defined as ${\mathcal{S}_p = \{ p \in H^0(\Omega) \}}$ and ${\mathcal{V}_p = \{ \vp \in H^0(\Omega) \}}$, where ${H^0(\Omega)}$ denotes the space of square integrable functions.
By multiplication of~\eqref{eq:momentum_equation} with~$\vu$ and~\eqref{eq:continuity_equation} with~$\vp$, integration over the domain~$\Omega$, and application of the divergence theorem, we obtain the weak form of the mechanical problem:
Find ${\vel \in \mathcal{S}_\vel}$ and ${p \in \mathcal{S}_p}$ such that
\begin{align} \label{eq:weakNavierStokes}
	\biggl( \rhoEff \left(\fracPartial{\vel}{t} + \left( \vel\cdot\nabla\right) \vel \right) , \vu \biggr)_{\Omega}
	- \biggl( p\, , \nabla\cdot\vu \biggr)_{\Omega}
	+ \biggl( 2\muEff\,\Bvep\, , \nabla\vu \biggr)_{\Omega}
	=& \biggl( \fRecoilPressureDiffuse + \fSurfaceTensionDiffuse + \fDarcy\, , \vu \biggr)_{\Omega} \\
	\biggl( \nabla\cdot\vel\, , \vp \biggr)_{\Omega} =&\,0 \\
	&\forall\, \{\vu, \vp\} \in \mathcal{V}_\vel \times \mathcal{V}_p \nonumber
\end{align}
holds.
Here, the operator ${\left( A, B \right)_X = \int_X A \cdot B\,\diffd \Bx}$ is the $\Ltwo$-scalar product over the domain~$X$.
Note that no advection stabilization, such as a streamline upwind Petrov--Galerkin (SUPG) term, is considered, since we assume that Reynolds numbers are moderate and element Péclet numbers are sufficiently small, cf.~\cite{kronbichler2018fast}.

\subsubsection{Heat equation}
\label{sec:cutfem_heat_transfer_model}

The strong form of the heat equation~\eqref{eq:heat_equation_two_phase} defines the heat transfer for the metal phase~$\OmegaL$ and the ambient gas phase~$\OmegaG$ individually.
The solution space for the metal temperature~$\Tl$ is restricted to finite elements that contain parts of the metal phase~$\OmegaL$, and, in variant~1, the same holds for the ambient gas temperature~$\Tg$ in~$\OmegaG$.
Finite elements cut by the metal--gas interface~$\Gamma$ are referred to as intersected elements and provide shape functions for~$\Tl$ that extend into~$\OmegaG$; in variant~1, they also provide shape functions for~$\Tg$, leading to overlapping solution spaces, where temperature values in the domain of the other phase appear as ghost values used solely for ghost-penalty stabilization~\cite{burman2010ghost}.
The interaction between the two phases is defined by the interface coupling conditions~\eqref{eq:interface_matching_two_phase} and~\eqref{eq:interface_matching_heat_flux_two_phase} at the metal--gas interface~$\Gamma$, embedded in an Eulerian domain.
In the following, we introduce a CutFEM~\cite{hansbo2002unfitted} approach for solving the sharp-interface two-phase heat equation.
We present two variants:
Variant~1 considers the coupled heat transfer in both the metal and ambient gas phases.
Variant~2 is a reduced and thus computationally more efficient approach, where the thermal problem is only solved in the metal domain, replacing the ambient gas phase with appropriate boundary conditions.
This simplification is justified because vapor flow is absent in melt pool models with unresolved evaporation, the process gas (e.g., argon) contributes negligibly to heat transfer due to its low thermal conductivity and heat capacity compared to molten metal, and the key melt pool phenomena occur within the molten metal or at the melt--gas interface.
However, variant~2 is unsuitable if vapor plume interactions are considered, e.g., through resolved evaporation models.

\paragraph{Variant~1: coupled two-phase heat transfer (\mbox{SI-2P})}
\label{sec:two_phase_description}
\!
\newline\indent
Variant~1 presents a sharp-interface CutFEM model for solving the coupled heat equation~\eqref{eq:heat_equation_two_phase} in the metal and ambient gas phases, considering the interface coupling conditions \eqref{eq:interface_matching_two_phase} and~\eqref{eq:interface_matching_heat_flux_two_phase}.
We denote this variant as the sharp-interface, two-phase (\mbox{SI-2P}) variant.
Each phase carries its own temperature field, i.e., the metal phase temperature~$\Tl$ in~$\OmegaL$ and the ambient gas phase temperature~$\Tg$ in~$\OmegaG$, as shown in the bottom right panel of Fig.~\ref{fig:hybrid_melt_pool_sketch}.
The solution spaces for the temperature fields are defined as ${\mathcal{S}_{\Tl} = \{ \Tl \in H^0(\OmegaL)\, |\, \Tl = \hat{T} \text{ on } \GammaD \}}$ and ${\mathcal{S}_{\Tg} = \{ \Tg \in H^0(\OmegaG)\, |\, \Tg = \hat{T} \text{ on } \GammaD \}}$, considering the Dirichlet boundary condition~\eqref{eq:Dirichlet_two_phase}.
The solution spaces for the corresponding test functions are defined as ${\mathcal{V}_{\Tl} = \{ \vl \in H^0(\OmegaL)\, |\, \vl = 0 \text{ on } \GammaD \}}$ and ${\mathcal{S}_{\Tg} = \{ \Tg \in H^0(\OmegaG)\, |\, \vg = 0 \text{ on } \GammaD \}}$.
To obtain the weak form of the heat equation, we multiply~\eqref{eq:heat_equation_two_phase} by the test functions~$\vl$ and~$\vg$, integrate over the respective domains, and apply the divergence theorem.
For the Neumann boundary condition, we insert~\eqref{eq:outer_Neumann_two_phase} in the boundary integral in~$\GammaN$.
The interface matching condition~\eqref{eq:interface_matching_two_phase} is enforced by a Nitsche term according to~\cite{ludvigsson2018high}, and the heat flux jump condition~\eqref{eq:interface_matching_heat_flux_two_phase} is applied to each phase's interface integral proportional to its relative thermal conductivity.
Including face-based ghost-penalty stabilization~\cite{burman2010ghost}, the weak form of the heat equation reads as follows:
Find ${\Tl \in \mathcal{S}_{\Tl}}$ and ${\Tg \in \mathcal{S}_{\Tg}}$ such that
\begin{align} \label{eq:weak_heat_equation_two_phase}
	0 =
	&\sum_{\phaseIndex \in \metalAndGas} \Biggl[\,
	\underbrace{\biggl( \rhoS\cpS \left( \frac{\partial \Ts}{\partial t} + \vel \cdot \nabla \Ts \right) , \vS \biggr)_{\OmegaS}
	+ \biggl( \condS \, \nabla \Ts \, , \nabla \vS \biggr)_{\OmegaS}}_{\text{convective and conductive heat transfer}}
	- \underbrace{\biggl( \qHat \, , \vS \biggr)_{\GammaN \cap \OmegaS}}_{\text{boundary heat flux}}
	\,\Biggr]
	\nonumber \\
	&- \underbrace{\biggl( \left\{ -\cond \, \partial_{\interfaceNormal} T \right\} , \left[ \test \right]_\Gamma \biggr)_\Gamma}_{\text{interface heat flux}}
	+ \underbrace{\biggl( \frac{\NitscheParamInterm}{h} \left[ T \right]_\Gamma, \left[ \test \right]_\Gamma \biggr)_\Gamma}_{\text{Nitsche term}}
	- \underbrace{\biggl( \left[ T \right]_\Gamma, \left\{ -\cond \, \partial_{\interfaceNormal} \test \right\} \biggr)_\Gamma}_{\text{symmetry term}}
	\nonumber \\
	&- \underbrace{\biggl( \convconbParamG \left( \qLaser + \qVapor(\Tl) \right) , \vl \biggr)_\Gamma
	- \biggl( \convconbParamL \left( \qLaser + \qVapor(\Tg) \right) , \vg \biggr)_\Gamma}_{\text{interface heat flux jump}}
	\nonumber \\
	&+ \underbrace{\sum_{\phaseIndex \in \metalAndGas} \left[
	\GPstabilM \, \GPstabilFunctionS ( \frac{\partial \Ts}{\partial t}, \vS )
	+ \frac{\GPstabilA}{h^2} \condS \, \GPstabilFunctionS ( \Ts, \vS )
	\right]}_{\text{ghost-penalty stabilization}} \\
	&\hspace{9.5cm}\forall\, \{\vl, \vg\} \in \mathcal{V}_{\Tl} \times \mathcal{V}_{\Tg} \nonumber
\end{align}
holds.
The expression~$\left\{ \bullet \right\}$ denotes a convex combination ${\left\{ x \right\} = \convconbParamL \inLiquid{x}} + \convconbParamG \inGas{x}$ with ${\convconbParamL = \condG / (\condL + \condG)}$ and ${\convconbParamG = \condL / (\condL + \condG)}$~\cite{burman2015cutfem}.
Here, ${\convconbParamG + \convconbParamL = 1}$ must hold to maintain unity.
In~\eqref{eq:weak_heat_equation_two_phase}, the summands in the first line represent the standard heat equation for the individual phases, including \emph{convective and conductive heat transfer}.
The \emph{boundary heat flux} term describes the Neumann boundary condition~\eqref{eq:outer_Neumann_two_phase} along the (fitted) domain boundary~$\GammaN$, while the \emph{interface heat flux} term describes the heat transfer between the two phases at the (embedded) interface~$\Gamma$.
With the \emph{Nitsche term}, the Dirichlet-type interface matching condition~\eqref{eq:interface_matching_two_phase} is enforced, considering Nitsche's method~\cite{nitsche1971uber}.
The variable~$\NitscheParamInterm$ is determined according to Burman et al.~\cite{burman2015cutfem} as ${\NitscheParamInterm = \NitscheParam\,\condG\,\condL / (\condG + \condL)}$, where~$\NitscheParam$ is the Nitsche parameter.
The \emph{symmetry term} restores the symmetry of the weak formulation.
Together with the Nitsche term, it maintains the coercivity of the weak formulation~\cite{benzaken2024contructing}.
In practical simulations, the optimal value of the Nitsche parameter~$\NitscheParam$ can be determined by initially choosing a relatively large value and then gradually reducing it until the limit of coercivity is reached.
For the simulation cases discussed in Section~\ref{sec:numerical_examples}, we found that a Nitsche parameter of ${\NitscheParam = 500}$ was sufficient to ensure coercivity.
The laser heat source and evaporation-induced cooling are imposed at the metal--gas interface $\Gamma$ as the \emph{interface heat flux jump}~\eqref{eq:interface_matching_heat_flux_two_phase}.
With the face-based \emph{ghost-penalty stabilization}~\cite{burman2010ghost}, we address the ill-conditioned linear system of equations for the case of very small cut finite elements, where a detailed description is given by \ref{sec:ghost_penalty}, considering the definition of the function~${\GPstabilFunctionS(\bullet, \vS)}$ according to~\eqref{eq:stabilization_function_j}.
The scalar constants~$\GPstabilM$ and~$\GPstabilA$ control the stabilization and influence the condition number of the linearized system of equations.
Their values are chosen to be ${\GPstabilM = 0.75}$ and ${\GPstabilA = 1.5}$, following the investigations by Sticko and Kreiss~\cite{sticko2016stabilized} and Burman and Hansbo~\cite{burman2012fictitious}, respectively.
In time-stepping schemes, ghost-penalty terms are only considered for implicit time evaluations.
\pagebreak

\paragraph{Variant~2: single-phase heat transfer with an immersed boundary (\mbox{SI-1P})}
\label{sec:one_phase_description}
\!
\newline\indent
Based on the assumption that heat transfer in the ambient gas phase can be neglected for melt pool models with unresolved evaporation, variant~2 presents a single-phase sharp-interface heat transfer model for the metal phase, denoted as the sharp-interface, one-phase (\mbox{SI-1P}) variant.
To that, we utilize an immersed boundary method~\cite{lee2003immersed} for the metal--gas interface, considering a CutFEM approach.
For this purpose, we consider the heat transfer equation~\eqref{eq:heat_equation_two_phase} and its initial and boundary conditions~\mbox{\eqref{eq:inital_condition_two_phase}-\eqref{eq:outer_Neumann_two_phase}} only for the metal domain~$\OmegaL$, and thus the unknown primary variable is the metal phase temperature $\Tl$.
At the metal--gas interface~$\Gamma$, a Neumann boundary condition is imposed
\begin{align} \label{eq:immersed_heat_flux_one_phase}
	-\condL \, \partial_{\interfaceNormal} T
	= \qGamma(\TGamma)
	\quad \text{on } \Gamma \times [0,t]
\end{align}
with the outward-pointing unit normal vector~$\interfaceNormal$ on~$\Gamma$, and the external heat flux~$\qGamma$ according to~\eqref{eq:laser_plus_evapor_heat_loss}.
Thus, we treat the metal--gas interface as adiabatic except for the externally applied heat flux~$\qGamma$.
To arrive at the weak formulation for the CutFEM approach, we follow the procedure outlined in Section~\ref{sec:two_phase_description} for \mbox{SI-2P}, but restrict the domain to the metal phase~$\OmegaL$.
The weak form for \mbox{SI-1P} reads as follows:
Find ${\Tl \in \mathcal{S}_{\Tl}}$ such that
\begin{align} \label{eq:weak_heat_equation_one_phase}
	0 =
	&\underbrace{\biggl( \rhoL\cpL \left( \frac{\partial \Tl}{\partial t} + \vel \cdot \nabla \Tl \right) , \vl \biggr)_{\OmegaL}
	+ \biggl( \condL \, \nabla \Tl \, , \nabla \vl \biggr)_{\OmegaL}}_{\text{convective and conductive heat transfer}}
	- \underbrace{\biggl( \qHat \, , \vl \biggr)_{\GammaN \cap \OmegaL}}_{\text{boundary heat flux}}
	\nonumber \\
	&- \underbrace{\biggl( \qLaser + \qVapor(\Tl) , \vl \biggr)_\Gamma}_{\text{interface heat flux}}
	+ \underbrace{\GPstabilM \, \GPstabilFunctionL ( \frac{\partial \Tl}{\partial t}, \vl )
	+ \frac{\GPstabilA}{h^2} \condL \, \GPstabilFunctionL ( \Tl, \vl )\biggl.}_{\text{ghost-penalty stabilization}} \\
	&\hspace{10cm}\forall\, \vl \in \mathcal{V}_{\Tl} \nonumber
\end{align}
holds.
Here, the first line comprises the standard heat equation, including \emph{convective and conductive heat transfer}, and the \emph{boundary heat flux} term that considers the Neumann boundary condition~\eqref{eq:outer_Neumann_two_phase} at the domain boundary~$\GammaN$.
The second line includes the boundary condition~\eqref{eq:immersed_heat_flux_one_phase} at the (unfitted) metal surface~$\Gamma$ through a prescribed \emph{interface heat flux}, and the face-based ghost-penalty stabilization~\cite{burman2010ghost} to address the ill-conditioned linear system of equations for very small cut finite elements, considering the function~$\GPstabilFunctionL(\bullet, \vl)$ according to~\eqref{eq:stabilization_function_j}.
Compared to \mbox{SI-2P}~\eqref{eq:weak_heat_equation_two_phase}, \mbox{SI-1P} does not require a Nitsche term, since no Dirichlet-type interface matching condition needs to be weakly enforced in the one-phase formulation.

\subsection{Extension of the interface temperature in a narrow band around the interface}
\label{sec:extension_of_interface_temperature}

In the DI two-phase flow model, regularized interface force distributions are spread across a finite transition layer of thickness~$\interfaceThickness$ centred around the sharp metal--gas interface~$\Gamma$.
The governing interface forces, namely recoil pressure and surface tension, depend on the interface temperature~$\TGamma$, which must therefore be known not only exactly on~$\Gamma$ but also throughout the associated narrow-band region where the regularized forces have support.
Consequently, an algorithm is required to reconstruct~$\TGamma$ in a narrow band around the sharp interface based on the CutFEM temperature field, so that the resulting interface temperature field can be consistently supplied to the DI two-phase flow model.
Furthermore, for the reduced but computationally efficient heat transfer model (\mbox{SI-1P}, described in Section~\ref{sec:one_phase_description}), the temperature is computed solely in the metal phase.
Thus, the local temperature in the diffuse interface region outside the metal phase is not available.

\begin{figure}[tb!]
	\centering
	\includegraphics[trim={0 0 3.8cm 0},clip]{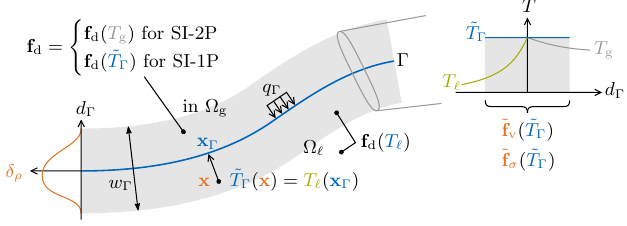}
	\caption{
		Sketch of the extended interface temperature and thereof dependent regularized interface forces in the diffuse transition region, which comprises a narrow band centred around the metal--gas interface~$\Gamma$ with a thickness of~$\interfaceThickness$.
		The extended interface temperature~$\TGammaExt$ is defined in the narrow band and at each point~$\Bx$ corresponds to the metal temperature~$\Tl$ value of the closest point on the interface~$\Bx_\Gamma$.
		The regularized force distributions, i.e., the evaporation-induced recoil pressure force~$\fRecoilPressureDiffuse$~\eqref{eq:recoil_pressure_diffuse} and the surface tension force~$\fSurfaceTensionDiffuse$~\eqref{eq:surface_tension_diffuse}, are computed based on the extended interface temperature~$\TGammaExt$.
		In the metal phase, the Darcy damping force~$\fDarcy$~\eqref{eq:darcy_damping_term} is computed based on the local temperature~$\Tl$.
		In the ambient gas phase~$\OmegaG$, the Darcy damping force~$\fDarcy$~\eqref{eq:darcy_damping_term} is evaluated differently for the two SI heat transfer variants (Section~\ref{sec:cutfem_heat_transfer_model}): For \mbox{SI-2P} it is based on the ambient gas temperature~$\Tg$, whereas for \mbox{SI-1P} it is based on the extended interface temperature~$\TGammaExt$, obtained by extending the metal interface temperature from~$\Gamma$ into the fictitious gas domain.
	}
	\label{fig:interface_temperature}
\end{figure}

Therefore, for the HSDI approach, we propose to extend the interface temperature~$\TGamma$, obtained from the CutFEM heat transfer model, in a narrow band around the interface, i.e, the diffuse transition region.
Fig.~\ref{fig:interface_temperature} provides a sketch of the diffuse transition region and the extension of the interface temperature.
Based on the extended interface temperature~$\TGammaExt$, temperature-dependent regularized interface forces are computed.
In this work, we consider an interface extension procedure based on the closest-point projection, which builds on the algorithm described in \mbox{\cite{schreter2024consistent, schreter2023evaluation}}.
A careful comparison in terms of accuracy and performance with alternative projection schemes, such as the $\Ltwo$-projection, is planned for future research.
The key steps of temperature field extension are as follows:
First, we discretize the melt pool surface as a set of interface points ${\Bx_\Gamma \in X_\Gamma}$ using the Marching Cubes Algorithm~\cite{lorensen1987marching}.
Second, we process all nodes of the fluid mesh within a narrow band around the interface:
For each node~$\Bx$, we perform a closest-point search to identify the closest point $\Bx_\Gamma$ from the set of discrete interface points~$X_\Gamma$.
Third, we evaluate the interface temperature~$\TGamma$ at the closest point in the narrow band and assign it to the corresponding value of the extended temperature field, i.e., ${\TGammaExt(\Bx) = \Tl(\Bx_\Gamma(\Bx))}$ for both variants.

\begin{remark}
In most DI formulations, the interface temperature in the narrow band is approximated by the local temperature, i.e. ${\TGamma \approx T}$.
However, as shown in our recent contribution~\cite{much2024improved}, even for diffuse models, a higher accuracy is obtained when $\TGamma$ is computed by extending the temperature from the closest point on the interface~$\Gamma$, i.e., the zero-level-set isosurface.
\end{remark}

\subsection{Computation of temperature-dependent forces in a regularized interface}
\label{sec:diffuse_two_phase_flow_model}

The momentum balance equation~\eqref{eq:momentum_equation} includes three volume forces:
The evaporation-induced recoil pressure force~$\fRecoilPressureDiffuse$ and the temperature-dependent surface tension force~$\fSurfaceTensionDiffuse$, including thermal Marangoni effects, read
\begin{align}
	\fRecoilPressureDiffuse(\TGammaExt, \indicator) &= \pv(\TGammaExt) \, \normalDiffuse \, \deltaDensity(\indicator),
	\label{eq:recoil_pressure_diffuse} \\
	\fSurfaceTensionDiffuse(\TGammaExt, \indicator) &=
	\left( \surfTensCoeff(\TGammaExt)\curvatureDiffuse\normalDiffuse + \left( \identity - \normalDiffuse \otimes \normalDiffuse \right) \nabla\surfTensCoeff(\TGammaExt) \right) \deltaDensity(\indicator)
	\label{eq:surface_tension_diffuse}
\end{align}
with the second-order identity tensor~$\identity$ and the temperature-dependent surface tension coefficient~$\surfTensCoeff$
\begin{align} \label{eq:surface_tension_coeff}
	\surfTensCoeff(\TGammaExt) = \max\left( \surfTensCoeffRef + \surfTensCoeffTD(\TGammaExt - \TsurfTensRef), \surfTensCoeffRes \right).
\end{align}
Here, $\surfTensCoeffRef$ is the surface tension coefficient at its reference temperature~$\TsurfTensRef$, and $\surfTensCoeffTD$ is the temperature gradient of the surface tension.
With~\eqref{eq:surface_tension_coeff}, we ensure that the surface tension coefficient remains positive for arbitrary temperature values.
The interface forces~\eqref{eq:recoil_pressure_diffuse} and~\eqref{eq:surface_tension_diffuse} are represented as CSF distributions using a regularized approximation of the Dirac delta function, which is contained in a finite but small transition region~\cite{brackbill1992continuum}.
We employ a density-scaled CSF approach, employing the density-scaled delta function~$\deltaDensity$
\begin{align} \label{eq:delta_density_scaled}
	\deltaDensity(\indicator) = || \nabla \indicator ||_{2} \, \rhoEff(\indicator) \, \deltaDensityCcorr
	\quad \text{with} \quad
	\deltaDensityCcorr = \frac{2}{\rhoL + \rhoG},
\end{align}
which demonstrated an improved stability for surface tension force computations in two-phase flow~\cite{kothe1996volume}.
This approach ensures that the magnitude of the interface-force-induced acceleration is well-balanced across the interface~\cite{yokoi2014density}.
It has been used in melt pool simulations to model interface forces, such as recoil pressure and surface tension, e.g., by \mbox{\cite{zhu2021mixed, much2024improved, schreter2025consistent, yan2018fully}}.
The regularized interface normal~$\normalDiffuse$ and the mean curvature~$\curvatureDiffuse$ in the diffuse transition region are computed from a filtering procedure of the level-set function, as described in \mbox{\cite{zahedi2012spurious, schreter2024consistent}}.
Within the diffuse transition region, the regularized representations of the temperature-dependent interface forces in~\eqref{eq:recoil_pressure_diffuse} and~\eqref{eq:surface_tension_diffuse} are computed based on the extended interface temperature~$\TGammaExt$, as described in Section~\ref{sec:extension_of_interface_temperature} and shown in Fig.~\ref{fig:interface_temperature}.

The Darcy damping force~$\fDarcy$~\eqref{eq:darcy_damping_term} depends on the solid fraction~$\solidFraction$ that is computed from the local temperature according to~\eqref{eq:solid_fraction}.
For \mbox{SI-1P}, in the diffuse transition region, the local temperature in the diffuse interface region outside the metal phase is not computed.
In that case, we propose computing the Darcy damping force~\eqref{eq:darcy_damping_term} based on the extended interface temperature~$\TGammaExt$ in~$\OmegaG$ while using the local temperature~$\Tl$ in the metal domain as depicted in Fig.~\ref{fig:interface_temperature}.

\begin{remark}
Representing interface forces as CSF distributions within diffuse transition regions of finite thickness is not strictly conservative with respect to momentum unless additional constraints are imposed~\cite{francois2006balanced}.
We expect that computing interface-temperature-dependent CSF distributions based on the extended interface temperature does not introduce additional conservation errors.
\end{remark}

\begin{remark}
To obtain a well-resolved diffuse transition region, we set the interface thickness proportional to the finite element side length~$h$ so that the number of finite elements of the fluid mesh across the interface ${\numberOfElementsInInterface = \interfaceThickness / h}$ is at least 8, following recommendations of \mbox{\cite{much2024improved, olsson2007conservative}}.
In the case of adaptive mesh refinement, the finite element size remains constant throughout the diffuse transition region.
\end{remark}

\subsection{Numerical framework}
\label{sec:numerical_framework}

As the interface~$\Gamma$ moves, the set of intersected elements and thus the physically relevant solution space changes over time:
Elements leaving the physical domain are deactivated, while previously inactive elements may become intersected and newly activated, with no prior solution available.
To consistently initialize the temperature field in these newly activated elements, we employ the ghost-penalty extrapolation procedure of Schott~\cite{schott2017stabilized}, which extrapolates the previous solution onto the enlarged solution space by minimizing normal derivative jumps across faces between newly activated physical elements and their neighbouring elements.
A detailed description of this procedure for adapting to a new interface position is given in \ref{sec:ghost_penalty_extrapol}.

We employ a standard tensor-product based Gaussian numerical integration for non-cut finite elements and Saye's approach of dimension reduction of integrals~\cite{saye2015high} for intersected finite elements.
For the efficient computational evaluation of structured and unstructured quadrature, we employ the computational design presented in~\cite{bergbauer2025high}, utilizing the tensor-product structure of the underlying finite element shape functions on the hexahedral element, while working on a single point at a time.

The computation of the temperature-dependent regularized interface forces, as described in Section~\ref{sec:diffuse_two_phase_flow_model}, requires the evaluation of the extended interface temperature~$\TGammaExt$ in the diffuse interface transition region.
We pre-compute the extended interface temperature field~$\TGammaExt$ in the narrow band described in Section~\ref{sec:extension_of_interface_temperature} with an extension algorithm \mbox{\cite{schreter2024consistent, schreter2023evaluation}}.

For time integration, we employ (semi-)implicit time stepping schemes, along with a weakly partitioned operator-splitting approach, as outlined in Algorithm~\ref{algo:partitioned_scheme}.
\begin{algorithm}[tb!]
	\caption{Outline of the partitioned solution scheme for the multi-phase melt pool model with a HSDI approach.}\label{algo:partitioned_scheme}
	\small
	$\levelset \gets \levelset_0,\; \Tl \gets T_0 {\color{TUMGrau},\; (\Tg \gets T_0)},\; \vel \gets \vel_0,\; p \gets p_0$
	\Comment*[r]{set initial conditions}
	\While{$t < \tEnd$}
	{
		$t \gets t + \dt$ \;
		\Comment{step 1:~compute level-set and geometric quantities}
		$\levelset, \normalDiffuse , \curvatureDiffuse \gets \texttt{level\_set\_solver}(\levelset, \vel)$ (cf. algorithm in~\cite{schreter2024consistent}) \;
		$\indicator \gets \indicator(\levelset)$~\eqref{eq:indicator} \;
		\Comment{step 2:~adapt to new interface location}
		$\Tl{\color{TUMGrau}, (\Tg)} \gets \texttt{ghost\_penalty\_extrapolation}(\Tl{\color{TUMGrau}, (\Tg)})$ (cf. algorithm in~\cite{schott2017stabilized}) \;
		\Comment{step 3:~solve SI heat equation}
		$\Tl{\color{TUMGrau}, (\Tg)} \gets \texttt{heat\_transfer\_solver}(\Tl{\color{TUMGrau}, (\Tg)}, \vel)$ \;
		\Comment{step 4:~extend interface temperature}
		$\TGammaExt \gets \texttt{extension\_algorithm}(\Tl)$ (cf. algorithm in~\cite{schreter2024consistent,schreter2023evaluation}) \;
		\Comment{step 5:~solve incompressible Navier--Stokes equations}
		$\rhoEff \gets \rhoEff(\indicator)$~\eqref{eq:parameter_transition_arithmetic},
		$\; \muEff \gets \muEff(\indicator, T)$~(\eqref{eq:parameter_transition_arithmetic} and~\eqref{eq:metal_viscosity}) \;
		$\fRecoilPressureDiffuse \gets \fRecoilPressureDiffuse(\TGammaExt, \indicator, \normalDiffuse)$~\eqref{eq:recoil_pressure_diffuse},
		$\; \fSurfaceTensionDiffuse \gets \fSurfaceTensionDiffuse(\TGammaExt, \indicator, \normalDiffuse, \curvatureDiffuse)$~\eqref{eq:surface_tension_diffuse},
		$\; \fDarcy \gets \fDarcy(\vel, \Tl, (\TGammaExt), \indicator)$~\eqref{eq:darcy_damping_term} \;
		$\vel, p \gets \texttt{flow\_solver}(\vel, p, \rhoEff, \muEff, \fRecoilPressureDiffuse, \fSurfaceTensionDiffuse, \fDarcy)$ (cf. algorithm in~\cite{kronbichler2018fast}) \;
	}
\end{algorithm}
Each field is propagated implicitly within each framework, i.e., the heat transfer, flow, and level-set framework.
Coupling terms, such as convection velocity and temperature-dependent forces and physical properties, are treated explicitly, which introduces a time-step limit.
The overall time-step limit -- which is influenced not only by the capillary time-step limit~\cite{brackbill1992continuum} but also by the explicit coupling terms -- is empirically estimated in the following.
Time integration is performed using the Crank--Nicolson method for heat transfer, the \mbox{BDF-2} scheme for two-phase flow, and the backward Euler method for level-set advection.

To achieve high computational efficiency, we employ adaptive meshing schemes ensuring a high spatial resolution of the diffuse transition region.
Furthermore, we exploit highly efficient matrix-free operator evaluation~\cite{kronbichler2012generic} along with iterative solvers.
The overall framework is parallelized using an MPI-based domain decomposition approach, leveraging the infrastructure of the \texttt{deal.II} library~\cite{arndt2025deal} along with the incompressible Navier--Stokes solver \texttt{adaflo}~\cite{kronbichler2018fast}.

\section{Numerical examples}
\label{sec:numerical_examples}

The proposed approaches are investigated in the following using selected benchmark examples.
Starting with purely thermal test cases mimicking laser--metal interactions, the accuracy of the SI heat transfer model variants is compared to standard DI models.
In increasingly complex thermal two-phase flow problems, the HSDI approach is validated.
Finally, the capabilities and robustness of the presented methods are demonstrated in a practically relevant example resembling laser-based processing of metals.

\subsection{1D laser-induced heating of a static surface}
\label{sec:onedim_benchmark}

In this section, we assess the accuracy of the two variants of the SI heat transfer models, described in Sections~\ref{sec:two_phase_description} and~\ref{sec:one_phase_description}, compared to a standard DI heat transfer model presented in~\cite{much2024improved}.
To that end, we consider the benchmark example of \emph{laser-induced heating of a static surface}, introduced in~\cite{much2024improved}, which is a purely thermal problem mimicking laser--metal interactions.
We consider a 1D two-phase domain ${\Omega = \{x \in [-a, a] \}}$ with ${a = \SI{100}{\mu m}}$, which contains the metal phase~$\OmegaL$ and the ambient gas phase~$\OmegaG$, each occupying half of the domain, separated by the immobile interface at the centre.
A domain sketch is shown in the top right corner of Fig.~\ref{fig:onedim}.
The signed distance function is prescribed to ${\distance(x) = -x}$.
As no flow dynamics are considered (${\vel = \Bzero}$), convective heat transfer is omitted.
Typical material parameter values representative of \tiSixFour are employed and are listed in Table~\ref{tab:param_ti64}.
The initial temperature is uniform at ${T_0 = \SI{500}{K}}$.
The interface is subjected to the heat flux~$\qGamma$ comprising a constant laser heat flux ${\qLaser = \SI{e10}{W\per m^2}}$ and temperature-dependent evaporation-induced cooling~$\qVapor$, while the boundaries of the domain~($\GammaD$) are kept at ${\hat{T} = \SI{500}{K}}$.
For the subsequent convergence studies, the instationary temperature solution at ${t = \SI{e-5}{s}}$ is considered.
\begin{table}[tb!]
	\caption{
		Representative material parameters for \tiSixFour.
	}
	\label{tab:param_ti64}
	\renewcommand\cellset{\renewcommand\arraystretch{0.8}
		\setlength\extrarowheight{0mm}}
	\centering
	\small
	\begin{tabular}{lcccl}
		\toprule
		Parameter & Symbol & Value & Unit & Reference \\
		\midrule
		thermal conductivity metal & $\inMetal{\conductivity}$ & 28.63 & \si{W\per(m\,K)} &~\cite{boivineau2006thermophysical} \\
		thermal conductivity gas & $\inGas{\conductivity}$ & 0.02863 & \si{W\per(m\,K)} & factor of \num{e3} w.r.t.~$\inMetal{\conductivity}$ \\
		density metal & $\inMetal{\rho}$ & 4087 & \si{kg\per m^3} &~\cite{mohr2020precise} \\
		density gas & $\inGas{\rho}$ & 4.087 & \si{kg\per m^3} & factor of \num{e3} w.r.t.~$\inMetal{\rho}$ \\
		specific heat capacity metal & $\cpL$ & \num{1130} & \si{J\per(kg\,K)} &~\cite{boivineau2006thermophysical} \\
		specific heat capacity gas & $\cpG$ & \num{11.3} & \si{J\per(kg\,K)} & factor of \num{e2} w.r.t.~$\cpL$ \\
		dynamic viscosity melt & $\inMelt{\mu}$ & \num{3.5e-3} & \si{kg\per(m\,s)} &~\cite{mohr2020precise} \\
		dynamic viscosity gas & $\inGas{\mu}$ & \num{3.5e-4} & \si{kg\per(m\,s)} & factor of \num{e2} w.r.t.~$\inLiquid{\mu}$ \\
		dynamic viscosity solid & $\inSolid{\mu}$ & \num{0.35} & \si{kg\per(m\,s)} & numerical value \\
		\makecell[l]{surface tension at\\reference temperature} & $\surfTensCoeffRef$ & \num{1.493} & \si{N\per m} &~\cite{mohr2020precise} \\
		\makecell[l]{surface tension\\reference temperature} & $\TsurfTensRef$ & \num{1933} & \si{K} &~\cite{mohr2020precise} \\
		\makecell[l]{surface tension\\gradient coefficient} & $\surfTensCoeffTD$ & \num{1.9e-4} & \si{N\per(m\,K)} &~\cite{mohr2020precise} \\
		laser absorptivity & $\absorptivity$ & \num{0.35} & \dimless &~\cite{khairallah2016laser} \\
		boiling temperature & $\Tv$ & 3133 & \si{K} &~\cite{zhang2020element} \\
		latent heat of evaporation & $\hv$ & \num{8.84e6} & \si{J\per kg} & calculated using \\
		\makecell[l]{reference temperature for\\the sum of specific enthalpy} & $\Thvref$ & \num{538} & \si{K} & \makecell[l]{the latent heat of\\each component} \\
		molar mass & $M$ & \num{4.78e-2} & \si{kg\per mol} &~\cite{lin2023enhanced} \\
		sticking constant & $\cs$ & \num{1} & \dimless &~\cite{khairallah2016laser} \\
		liquidus temperature & $\Tliquidus$ & 2200 & \si{K} & numerical value \\
		solidus temperature & $\Tsolidus$ & 1933 & \si{K} & numerical value \\
		\makecell[l]{parameter that depends on\\the mushy zone morphology} & $\darcyMorthology$ & \num{e11} & \si{kg\per(m^3 s)} & numerical value \\
		\makecell[l]{parameter to avoid\\division by zero} & $\darcyDivZero$ & \num{1} & \dimless & numerical value \\
		\bottomrule
	\end{tabular}
\end{table}

For the two SI heat transfer model variants, we discretize each simulation with a uniform finite element mesh with an odd number of elements to ensure that the metal--gas interface~$\Gamma$ is located inside a finite element and does not coincide with a finite element edge.
A time step of ${\dt = \SI{e-9}{s}}$ is used for the Crank--Nicolson time integration, which showed a negligible temporal discretization error in a temporal convergence study.

For comparison, we consider a DI heat transfer model, in which the volume-specific heat capacity ${\cv = \rho\cp}$ is interpolated across the diffuse transition region via~\eqref{eq:parameter_transition_arithmetic}, and the parameter-scaled CSF approach~\cite{much2024improved} is chosen to be proportional to~$\cvEff$.
The temperature-dependent evaporation-induced cooling~$\qVapor$ is determined based on the interface temperature~$\TGamma$, corresponding to the best-performing variant in~\cite{much2024improved}.
For the spatial and temporal discretization, we choose a uniform finite element mesh and the backward Euler time integration scheme with a time step of ${\dt = \SI{e-9}{s}}$.
The interface thickness~$\interfaceThickness$ is chosen based on the number of finite elements across the diffuse transition region ${\numberOfElementsInInterface}$ and thus depends on the finite element size~$h$ by ${\interfaceThickness = \numberOfElementsInInterface h}$.
We will evaluate the temperature accuracy for different interface thickness values by varying ${\numberOfElementsInInterface\in\{8, 16, 32\}}$.

The numerical reference solutions~$\Tref$ are obtained using a fitted interface model, where the metal--gas interface~$\Gamma$ coincides with a finite element edge, applying the interface fluxes on that discrete edge.
A very fine spatial and temporal discretization is employed to ensure a converged solution, i.e., ${h = \SI{1.56e-3}{\mu m}}$ and ${\dt = \SI{e-10}{s}}$.
\mbox{SI-2P} and the DI heat transfer models are compared to a two-phase reference solution.
Accordingly, \mbox{SI-1P} is compared to a single-phase reference solution that only comprises the metal phase~$\OmegaL$ with a Neumann boundary condition ${-\condL\outerNormal T = \qGamma}$ at the interface position.

In the following, we will discuss two increasingly complex cases:
In Section~\ref{sec:onedim_no_evapor}, we neglect evaporation-induced cooling to simplify the problem, while in Section~\ref{sec:onedim_with_evapor}, we consider evaporation-induced cooling according to~\eqref{eq:evaporative_heat_loss_with_specific_enthalpy}.

\subsubsection{Without evaporation-induced cooling}
\label{sec:onedim_no_evapor}

\begin{figure}[tb!]
	\centering
	\includegraphics{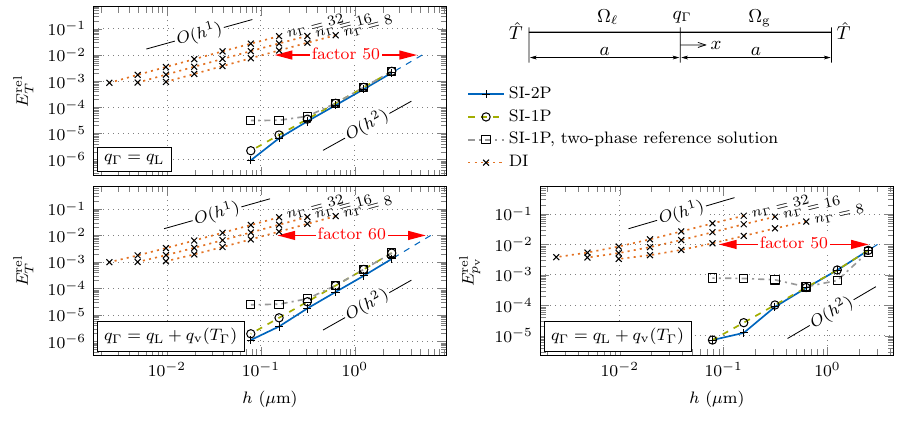}
	\caption{
		Error measures of the instationary temperature solution at ${t = \SI{e-5}{s}}$ for an example considering the laser-induced heating of a static surface benchmark~\cite{much2024improved}, including a sketch of the domain (top right).
		Comparison between \mbox{SI-2P}, \mbox{SI-1P}, and the DI heat transfer model~\cite{much2024improved} with a number of ${\numberOfElementsInInterface \in \{8,16,32\}}$ finite elements across the diffuse transition region: relative temperature error over finite element size~$h$
		(top~left)~without evaporation-induced cooling, i.e., Section~\ref{sec:onedim_no_evapor} and
		(bottom~left) with evaporation-induced cooling, i.e., Section~\ref{sec:onedim_with_evapor};
		(bottom~right)~relative error in the evaporation-induced recoil pressure over finite element size~$h$.
		The 1\% error level is highlighted.
	}
	\label{fig:onedim}
\end{figure}

For this section, the interface heat flux only contains the laser heat flux ${\qGamma = \qLaser}$, while the evaporation-induced cooling remains zero ${\qVapor = 0}$.
This simplification results in a linear system of equations for solving the heat equation.
In the top left panel of Fig.~\ref{fig:onedim}, the relative $\Ltwo$-error ${\TerrRel = ||T - \Tref||_{\Ltwo} / ||\Tref||_{\Ltwo}}$ of the instationary temperature field solution is shown over the finite element size~$h$ at ${t = \SI{e-5}{s}}$.
Both SI heat transfer model variants exhibit a similar convergence behaviour with a convergence order of~$O(h^2)$, while \mbox{SI-2P} achieves a slightly better accuracy at every refinement level.
Note that for \mbox{SI-1P}, i.e., the single-phase approach, the temperature field and thus the $\Ltwo$-norm are restricted to the metal domain~$\OmegaL$.
Additionally, Fig.~\ref{fig:onedim} shows the $\Ltwo$-error for \mbox{SI-1P} compared to the two-phase reference solution.
It exhibits asymptotic behaviour towards ${\TerrRel \approx \num{3e-5}}$, which represents the modelling error of \mbox{SI-1P}, as it neglects the heat transfer in the ambient gas phase.
The comparative DI heat transfer model only exhibits a convergence order of~$O(h^1)$, whereas the SI variants are at least two orders of magnitude more accurate for the same mesh.
In~\cite{much2024improved}, a benchmark tolerance level for DIDI approaches was specified to 1\% relative error.
To achieve the benchmark accuracy, the SI variants allow for a finite element size~$h$ that is almost two orders of magnitude larger than that of the DI approach.

\subsubsection{With evaporation-induced cooling}
\label{sec:onedim_with_evapor}

In this section, we investigate the \emph{laser-induced heating of a static surface} benchmark example, including the evaporation-induced cooling~$\qVapor$ according to~\eqref{eq:evaporative_heat_loss_with_specific_enthalpy}, i.e., the interface heat flux is defined as ${\qGamma = \qLaser + \qVapor(\TGamma)}$~\eqref{eq:laser_plus_evapor_heat_loss}.
Since evaporation-induced cooling commences above the boiling temperature, which is only reached towards the end of the time interval for this specific example, no significant difference is expected in the temperature profile compared to Section~\ref{sec:onedim_no_evapor}.
The bottom left panel of Fig.~\ref{fig:onedim} shows the relative $\Ltwo$-error of the instationary temperature field over the finite element size~$h$.
Similar to the results in Section~\ref{sec:onedim_no_evapor}, the two SI variants achieve a convergence order of~$O(h^2)$, while the DI heat transfer model shows only a convergence order of~$O(h^1)$.
\mbox{SI-2P} is slightly more accurate than \mbox{SI-1P} at every refinement level.
At the same finite element size~$h$, both SI variants are at least two orders of magnitude more accurate than the DI model.

In addition to the temperature field, we analyze the recoil pressure error, as the recoil pressure~$\pv$ is the dominant driving force in melt pool dynamics and is highly sensitive to the interface temperature.
The bottom right panel of Fig.~\ref{fig:onedim} shows the relative error ${\pvErrRel = | \pv(\TGamma) - \pvRef | / \pvRef}$ in the recoil pressure at the interface, which is computed from the instationary temperature solution at the sharp metal--gas interface ${\Gamma = \{x = 0\}}$, i.e., ${\pv(\TGamma) = \pv(T(x = 0))}$ and ${\pvRef(\TGammaRef) = \pv(\Tref(x = 0))}$.
The SI approaches clearly outperform the DI model, considering the accuracy at the same finite element size~$h$.
The $\Ltwo$-error for \mbox{SI-1P} compared to the two-phase reference solution exhibits asymptotic behaviour towards ${\pvErrRel \approx \num{8e-4}}$, representing its modelling error.
To reach the benchmark tolerance level of 1\%, the DI model requires a very narrow interface thickness and a very fine mesh resolution~\cite{much2024improved} (${h = \SI{0.07}{\mu m}}$ for ${\numberOfElementsInInterface = 8}$), while the SI approach achieves that tolerance with an extrapolated finite element side length of ${h \approx \SI{3}{\mu m}}$, meaning the mesh can be about \num{50} times coarser.

In conclusion, for a given mesh, the SI heat transfer model variants presented in Sections~\ref{sec:two_phase_description} and~\ref{sec:one_phase_description} achieve significantly higher accuracy regarding temperature and, critically, recoil pressure (at least two orders of magnitude) than the best-performing DI model variant presented in~\cite{much2024improved}.
To achieve the 1\% error bound, the mesh resolution of the SI model is allowed to be one to two orders of magnitude coarser.

\subsection{2D laser-induced heating of a fixed melt pool surface}
\label{sec:fixed_benchmark}

\begin{figure}[tb!]
	\centering
	\includegraphics{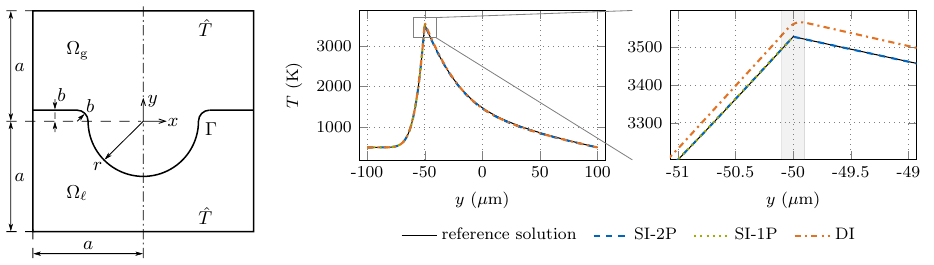}
	\caption{
		Laser-induced heating of a 2D fixed melt pool surface benchmark example:
		(left)~Sketch of the domain and interface geometry~\cite{much2024improved}, with ${a = \SI{100}{\mu m}}$, ${r = \SI{50}{\mu m}}$, and ${b = \SI{10}{\mu m}}$;
		(right)~instationary temperature field at ${t = \SI{e-5}{s}}$ along the $y$-axis.
		Comparison between \mbox{SI-2P}, \mbox{SI-1P}, the DI heat transfer model~\cite{much2024improved}, and the reference solution.
		The interface thickness~$\interfaceThickness$ of the diffuse transition region for the diffuse-interface model is highlighted.
	}
	\label{fig:fixed}
\end{figure}

As a second test case, we consider the \emph{laser-induced heating of a fixed melt pool surface} benchmark example, which was introduced in~\cite{much2024improved} to study DI approaches.
In this purely thermal problem, the flow dynamics are not resolved (${\vel = \Bzero}$), thereby omitting convective heat transfer.
With a spatially fixed interface geometry mimicking a laser-induced melt pool depression, this problem is used to assess the accuracy of temperature-dependent interface flux modelling on a curved interface.
We consider the 2D domain ${\Omega = \{\Bx \in [-a, a]^2\}}$ with the length parameter ${a = \SI{100}{\mu m}}$, which is occupied by the metal phase~$\OmegaL$ and the ambient gas phase~$\OmegaG$ that are separated by the metal--gas interface~$\Gamma$.
The fixed interface geometry is shown in the left panel of Fig.~\ref{fig:fixed} and characterized by the radii ${r = \SI{50}{\mu m}}$ and ${b = \SI{10}{\mu m}}$.
The signed distance~$\distance$ to the metal--gas interface~$\Gamma$ is prescribed to
\begin{align*}
	\distance(\Bx) =
	\begin{cases}
		||\Bx||-r
		&\text{for} \quad |x| < r+b \; \wedge \; y < 0 \\
		\min\left(||\Bx||-r, b-y\right)
		&\text{for} \quad |x| \geq r+b \; \wedge \; y < 0 \\
		b-y
		&\text{for} \quad |x| \geq r+b \; \wedge \; y \geq 0 \\
		b-\sqrt{(r+b - |x|)^2 + y^2}
		&\text{for} \quad |x| < r+b \; \wedge \; y \geq 0 \\
	\end{cases}.
\end{align*}
The temperature is initially uniform at ${T_0 = \SI{500}{K}}$ and prescribed to ${\hat{T} = \SI{500}{K}}$ at the top and bottom boundaries~($\GammaD$), while the left and right boundaries~($\GammaN$) are adiabatic (${\qHat = 0}$ in~\eqref{eq:outer_Neumann_two_phase}).
The metal--gas interface is subject to the laser heat flux~$\qLaser$ and the evaporation-induced cooling~$\qVapor$ according to~\eqref{eq:laser_plus_evapor_heat_loss}, where the laser heat source is defined as
\begin{align} \label{eq:laser_gauss_sharp}
	\qLaser(\Bx) = \absorptivity\,\laserPower\,\frac{2}{\laserRadius^2 \pi}\,\langle\interfaceNormal\laserDirection\rangle \exp\left(-2\left(\frac{\ofLaser{d}(\Bx)}{\laserRadius}\right)^2\right) \quad \text{on } \Gamma
\end{align}
and models a spatially fixed laser with a Gaussian profile~\cite{meier2021novel} with the absorptivity~$\absorptivity$, the laser power ${\laserPower = \SI{250}{W}}$, and the laser radius ${\laserRadius = \SI{70}{\mu m}}$.
Here, $\ofLaser{d}(\Bx)$ is the distance between and the laser beam centre line, defined by the laser position ${\laserPosition = \Bzero}$ and the laser direction ${\laserDirection = -\Be_y}$ corresponding to the negative $y$-direction, and a point~$\Bx$.
The Macauley bracket~$\langle\bullet\rangle$ returns the input value if positive, and zero otherwise.
Representative material parameter values for \tiSixFour are employed, as listed in Table~\ref{tab:param_ti64}.
We investigate the instationary temperature solution at ${t = \SI{e-5}{s}}$.

For both SI heat transfer model variants, a uniform Cartesian mesh consisting of quadrilateral finite elements with an element side length of ${h = \SI{0.391}{\mu m}}$ and the time step size ${\dt = \SI{e-9}{s}}$ are employed.
For comparison, we consider the results from~\cite{much2024improved}, which were obtained using a DI heat transfer model which employs a parameter-scaled delta CSF approach and computes temperature-dependent interface flux distributions based on the extended interface temperature.
We consider the results of the finest spatial resolution in~\cite{much2024improved}, i.e., an interface thickness of ${\interfaceThickness = \SI{0.2}{\mu m}}$ and a Cartesian finite element mesh of bilinear quadrilateral elements that is locally refined to a minimum finite element side length of ${h = \SI{0.0061}{\mu m}}$ in the diffuse transition region.
The reference solution~$\Tref$ was obtained in~\cite{much2024improved} by a fitted interface model using an aligned mesh whereby the metal--gas interface~$\Gamma$ coincides with finite element edges.
A very fine spatial and temporal discretization with ${h = \SI{0.047}{\mu m}}$ at the interface and ${\dt = \SI{e-10}{s}}$ ensures a converged solution.

The right panel of Fig.~\ref{fig:fixed} shows the instationary temperature fields at ${t = \SI{e-5}{s}}$ along the \mbox{$y$-axis} for the two SI variants and the comparative model compared to the reference solution.
While the two variants of the SI heat transfer models are in good agreement with the reference solution, the comparative model deviates visibly in the close vicinity of the metal--gas interface~$\Gamma$.
Table~\ref{tab:fixed_accuracy} lists various quantities of interest and error measures for comparing the quality of the proposed SI models to the DI model.
\begin{table}[tb!]
	\caption{
		Laser-induced heating of a 2D fixed melt pool surface benchmark example:
		comparison between the SI heat transfer model variants, and the DI model~\cite{much2024improved}.
		The relative maximum temperature error is computed by $\TmaxErrRel = |\Tmax-\Tmaxref| / \Tmaxref$ with the maximum reference temperature~$\Tmaxref = \SI{3529.2}{K}$, and the relative maximum recoil pressure error is computed by ${\pvMaxErrRel = |\pvMax-\pvMaxRef| / \pvMaxRef}$ with the maximum reference recoil pressure ${\pvMaxRef = \SI{332.2}{kPa}}$.
	}
	\label{tab:fixed_accuracy}
	\centering
	\small
	\begin{tabular}{lccc}
		\toprule
		& \mbox{SI-2P} & \mbox{SI-1P} & DI \cite{much2024improved} \\
		\midrule
		$h$ & \SI{0.391}{\mu m} & \SI{0.391}{\mu m} & \SI{0.0061}{\mu m} \\
		$\Tmax$ & \SI{3529.1}{K} & \SI{3529.3}{K} & \SI{3567.5}{K} \\
		$\TmaxErrRel$ & \num{0.0028}\% & \num{0.0028}\% & \num{1.1}\% \\
		$\pvMax$ & \SI{332.0}{kPa} & \SI{332.4}{kPa} & \SI{387.6}{kPa} \\
		$\pvMaxErrRel$ & \num{0.06}\% & \num{0.06}\% & \num{16.7}\% \\
		\bottomrule
	\end{tabular}
\end{table}
\mbox{SI-1P} achieves the same level of accuracy as the two-phase reference solution with respect to \mbox{SI-2P}.
This demonstrates that our model assumption, that the ambient gas phase can be neglected due to its low energy contribution to heat transfer, results in a negligible modelling error.
The DI model overestimates the interface temperature, resulting in a significant error (16.7\%) in the recoil pressure.
In addition, it has the finest spatial resolution with a finite element side length (${h = \SI{0.0061}{\mu m}}$) that is two orders of magnitude smaller than the other approaches and impractical for use in 3D simulations.

Both SI heat transfer model variants, presented in Sections~\ref{sec:two_phase_description} and~\ref{sec:one_phase_description}, achieve a significantly better accuracy compared to the DI model -- even at significantly coarser spatial discretizations.
By accurately representing discontinuities at the interface, SI approaches not only improve the accuracy of predicting the interface temperature but also enable accurate simulations at spatial resolutions suitable for practical 3D applications.

\subsection{Thermo-capillary droplet migration}
\label{sec:thermo_capillary_droplet}

Before analyzing the proposed HSDI approach in complex melt pool dynamics, we perform a verification for general thermal two-phase flows, including temperature-dependent Marangoni effects, based on the \emph{thermo-capillary droplet migration} studied by Ma and Bothe~\cite{ma2011direct}.
This benchmark provides a valuable test for consistency, efficiency, and robustness, even though only minor accuracy gains are
expected due to the small interfacial thermal gradients in this example.
Thus, consideration of a laser heat source, evaporation effects, as well as melting and solidification, is postponed to the subsequent section.

A circular droplet with a radius of ${a = \SI{1.44e-3}{m}}$ is initially located at the centre of a square domain with a side length of $4a$, i.e., ${\Omega = \{ \Bx \in [-2a,2a]^2 \}}$.
The initial signed distance is $\distance(\Bx) = a - ||\Bx||$.
At the bottom boundary (${y = -2a}$), the temperature is prescribed to ${\hat{T}_1 = \SI{290}{K}}$ and at the top boundary (${y = 2a}$), it is prescribed to ${\hat{T}_2 = \SI{291.152}{K}}$, while the lateral boundaries are considered adiabatic.
A uniform gradient of ${||\nabla T_0|| = \SI{200}{K\per m}}$ defines the initial temperature field.
For the velocity, no-slip conditions~\eqref{eq:no_slip_BC} are applied to the top and bottom boundaries and slip conditions~\eqref{eq:slip_BC} to the lateral boundaries, and the initial velocity is at rest ${\vel_0 = \Bzero}$.
The material parameters of the ambient phase~$\OmegaG$ are chosen as ${\condG = \SI{2.4e-6}{W\per(m\,K)}}$, ${\rhoG = \SI{500}{kg\per m^3}}$, ${\cpG = \SI{e-4}{J\per(kg\,K)}}$, and ${\inGas{\mu} = \SI{0.024}{kg\per(m\,s)}}$, while the droplet parameters in~$\OmegaL$ are taken as half of each of these values.
The temperature-dependent surface tension~\eqref{eq:surface_tension_diffuse} is defined by ${\surfTensCoeff = \SI{e-2}{N\per m}}$, ${\TsurfTensRef = \SI{290}{K}}$, and ${\surfTensCoeffTD = \SI{2e-3}{N\per(m\,K)}}$.
In terms of dimensionless numbers, this test case can be characterized by the Reynolds number ${\Re = \rhoG a u_r/ \inGas{\mu} = 0.72}$, the Marangoni number ${\Ma = \rhoG\cpG a u_r / \condG = 0.72}$, and the capillary number ${\Ca = \inGas{\mu} u_r / \surfTensCoeff = 0.0576}$, where the reference velocity is defined as ${u_{\*{r}} = \surfTensCoeffTD ||\nabla T_0|| a / \inGas{\mu} = \SI{0.024}{m\per s}}$.

For the heat transfer model, we consider the \mbox{SI-2P} approach, cf. Section~\ref{sec:two_phase_description}, and for the DI flow framework, we choose the interface thickness such that the number of finite elements across the interface is ${\numberOfElementsInInterface = 8}$.
We use the identical spatial discretization as in the reference solution~\cite{ma2011direct}, i.e., a Cartesian mesh of quadrilateral finite elements with a uniform element side length of ${h = a / 16 = \SI{0.09e-3}{m}}$, resulting in the interface thickness ${\interfaceThickness = \SI{0.72e-3}{m}}$.
A constant time step size of ${\dt = \SI{0.5e-3}{s}}$ is employed.

\begin{figure}[tb!]
	\centering
	\includegraphics{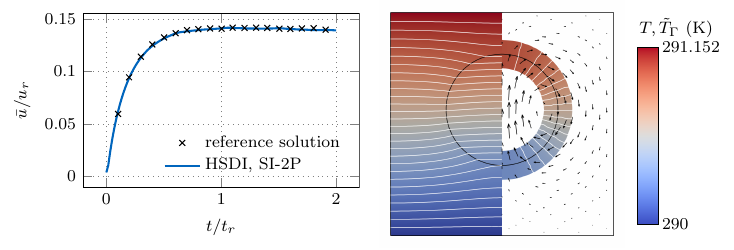}
	\caption{
		Thermo-capillary droplet migration:
		(left)~dimensionless velocity~$\bar{u}/u_r$ over dimensionless time~$t/t_r$ compared with the reference solution~\cite{ma2011direct};
		(right)~configuration at dimensionless time ${t / t_r = 2}$:
		The interface~$\Gamma$ is indicated by a black line.
		Left half: temperature field and its isothermal lines; right half: extended interface temperature~$\TGammaExt$ and its isothermal lines in the diffuse transition region, and arrows indicating the velocity field.
	}
	\label{fig:thermo_capillary_droplet}
\end{figure}

The left panel of Fig.~\ref{fig:thermo_capillary_droplet} shows the dimensionless upward velocity ${\bar{u} / u_r}$ over the dimensionless time ${t / t_r}$, where the reference time is ${t_r = a / u_r = \SI{0.06}{s}}$.
Here,~$\bar{u}$ is computed by numerical differentiation of the droplet centroid position with respect to time.
Due to the prevailing temperature gradient, the temperature-dependent surface tension~\eqref{eq:surface_tension_diffuse} causes Marangoni convection, leading to an upward droplet migration.
The HSDI approach agrees well with the reference solution.
In the right panel of Fig.~\ref{fig:thermo_capillary_droplet}, the temperature field and its isothermal lines are shown at dimensionless time ${t / t_r = 2}$.
The isothermal lines exhibit kinks at the interface~$\Gamma$, demonstrating the sharp-interface representation.
Additionally, the velocity field and the extended interface temperature~$\TGammaExt$, cf. Section~\ref{sec:extension_of_interface_temperature} is shown in the diffuse transition region.
The isotherms of the extended interface temperature field are perpendicular to the interface, demonstrating that the extension is accurately performed along the interface normal direction.
\pagebreak

\subsection{Benchmark: 2D laser-induced vapor depression}
\label{sec:angled}

In this section, we propose a \emph{laser-induced vapor depression} benchmark example for assessing the accuracy of thermal two-phase flow models in a setting representative of melt pool dynamics and tailored to test its specific features.
The setup includes the typically large contrast in physical properties between metal and ambient gas, a laser-induced heat flux at the metal surface, evaporation-induced recoil pressure and cooling, and temperature-independent surface tension.
We intentionally choose a 2D setup to allow for detailed convergence analyses, excluding the additional complexity of melting and solidification as well as the temperature dependence of surface tension and wetting forces.
This provides controlled and robust test cases that are independent of any particular melting and solidification or wetting-force model.
Furthermore, it enables the assessment of the HSDI approach compared to a purely DI melt pool model.
The analyzed metrics are vapor depression depth and interface geometry, which reflect practically relevant features of melt pool dynamics and can be readily verified.

A domain sketch is shown in the top left panel of Fig.~\ref{fig:angled}.
\sloppy{
We consider the 2D domain ${\Omega = \{\Bx \in [-a, a]^2\}}$, with the length parameter $a$, which contains the metal phase~$\OmegaL$ and the ambient gas phase~$\OmegaG$ separated by the metal--gas interface~$\Gamma$ that is defined by its initial signed distance function $\distance(\Bx) = -y$ in the initial state.
}
The initial temperature is uniform at ${T_0 = \SI{500}{K}}$, and the initial velocity is at rest ${\vel_0 = \Bzero}$.
At the top (${y = a}$) and bottom (${y = -a}$) boundaries, the temperature is prescribed to ${\hat{T} = \SI{500}{K}}$ while the lateral boundaries are considered adiabatic.
For the velocity, slip conditions~\eqref{eq:slip_BC} are considered on all boundaries.
At the metal--gas interface~$\Gamma$, we consider a laser heat flux~$\qLaser$ according to~\eqref{eq:laser_gauss_sharp} with the laser power~$\laserPower$, radius ${\laserRadius = \SI{70}{\mu m}}$, and position ${\laserPosition = \Bzero}$.
The laser direction ${\laserDirection = [-\sin{\alpha}, -\cos{\alpha}]^\top}$ is tilted by an angle of ${\alpha = 5^\circ}$ as shown in the top left panel of Fig.~\ref{fig:angled} to break symmetry.
For simplicity, we neglect melting and solidification effects, i.e., we assume a vanishing Darcy damping force ${\fDarcy = \Bzero}$ and the metal viscosity equals the melt viscosity ${\inLiquid{\mu} = \inMelt{\mu}}$ at any temperature.
Furthermore, we assume a constant surface tension coefficient ${\surfTensCoeff = \SI{1.493}{N\per m}}$ with ${\surfTensCoeffTD = 0}$, leading to vanishing Marangoni effects.
Apart from the abovementioned exceptions, we employ material parameter values representative of \tiSixFour, as listed in Table~\ref{tab:param_ti64}.

In the following, we consider the HSDI approach with the two SI heat transfer model variants presented in Section~\ref{sec:cutfem_heat_transfer_model}.
For the DI two-phase flow model, we choose the interface thickness~$\interfaceThickness$ to be proportional to the finite element side length~$h$ so that the number of elements across the diffuse transition region is ${\numberOfElementsInInterface = 16}$, aiming to minimize the spatial discretization error due to a poorly resolved diffuse transition region.

For comparison, we consider the purely DI melt pool thermo-hydrodynamics model from~\cite{much2024improved}.
In the DI heat transfer model, the volume-specific heat capacity ${\cv = \rho\cp}$ is interpolated across the diffuse transition region, and the parameter-scaled CSF approach~\cite{much2024improved} is proportional to~$\cv$.
The temperature-dependent evaporation-induced cooling~$\qVapor$ and recoil pressure~$\fRecoilPressureDiffuse$ are determined based on the extended interface temperature~$\TGammaExt$.
We choose the interface thickness ${\interfaceThickness = 16h}$ identical to the HSDI approach.
The underlying DI two-phase flow model is the same for the proposed HSDI and the purely DI thermal two-phase flow approach.
A constant time step size~$\dt$ is used for all simulations, as well as a Cartesian mesh of quadrilateral finite elements that is locally refined to a minimum element side length~$h$ in the vicinity of the diffuse transition region.

For the numerical reference solution, we consider the results from the HSDI approach with the \mbox{SI-2P} heat transfer model using a minimum finite element side length~$h$ and interface thickness~$\interfaceThickness$ that are four times smaller than the finest mesh of the simulations discussed in the following.

\subsubsection{Stable vapor depression}
\label{sec:angled_stable}

\begin{figure}[tb!]
	\centering
	\includegraphics{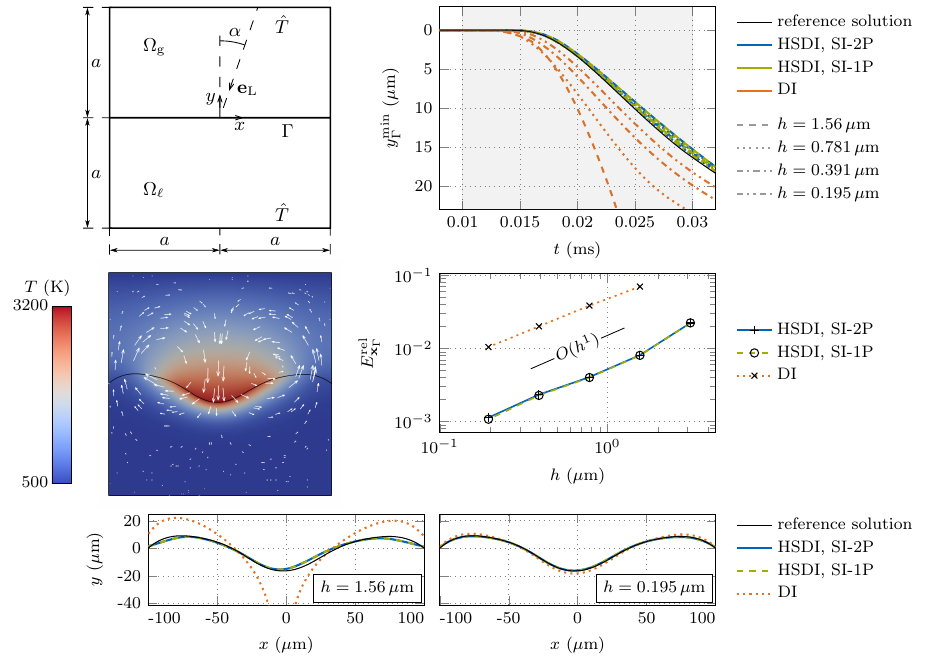}
	\caption{
		(top~left)~sketch of the domain, initial interface, and angled laser direction of the 2D vapor depression examples;
		stable vapor depression:
		(top~right)~vapor depression depth~$\interfaceDepthTwoD$ over time for different mesh resolutions.
		The time interval considered for the interface error norm~$\contourErrRel$ is highlighted.
		(centre~left)~Configuration of the reference solution at ${t = \SI{0.03}{ms}}$:
		The metal--gas interface~$\Gamma$ is indicated by a black line, and arrows indicate the velocity field.
		(centre~right)~Interface error norm~$\contourErrRel$~\eqref{eq:contour_error} over finite element size~$h$;
		(bottom)~metal--gas interface contours~$\Bx_\Gamma$ at ${t = \SI{0.03}{ms}}$ for ${h = \SI{1.56}{\mu m}}$ and ${h = \SI{0.195}{\mu m}}$.
		All plots compare the HSDI approach with \mbox{SI-2P}, \mbox{SI-1P}, and the purely DI melt pool model~\cite{much2024improved}.
	}
	\label{fig:angled}
\end{figure}

In this section, we focus on the initial stable formation of the vapor depression, before the high-frequency oscillations of the melt pool surface start to develop.
Thereto, we employ a length parameter of ${a = \SI{100}{\mu m}}$, a laser power of ${\laserPower = \SI{150}{W}}$, and a time step size of ${\dt = \SI{e-8}{s}}$.
The vapor depression becomes unstable after ${t = \SI{0.03}{ms}}$, so we consider the time interval up to that point.
The centre left panel of Fig.~\ref{fig:angled} shows the configuration of the reference solution at ${t = \SI{0.03}{ms}}$.

In the top right panel of Fig.~\ref{fig:angled}, the depth of the laser-induced vapor depression~$\interfaceDepthTwoD$ over time is shown for the HSDI approach variants and the comparative DI model for different minimum finite element side lengths~$h$.
The two HSDI heat transfer model variants yield practically identical results at every investigated finite element side length~$h$.
For the coarsest mesh, the HSDI approach is already close to the reference solution and exhibits convergence.
For the DI model, convergence towards the reference solution is noted; however, the deviations from the reference solution for the same finite element size are significantly larger compared to the HSDI approach.

To quantify the error in the temporal evolution of the metal--gas interface contour~$\Bx_\Gamma$, we employ a time-averaged interface error norm~$\contourErrRel$ according to
\begin{align} \label{eq:contour_error}
	\contourErrRel = \frac{
		\int_{t_1}^{t_2}
		\int_{-a}^{a}
		\left| y_\Gamma(x,t) - y_\Gamma^{\*{ref}}(x,t) \right|
		\diffd x\,\diffd t
	}{2ab(t_2 - t_1)},
\end{align}
where the deviation from the reference interface contour~$\Bx_\Gamma^{\*{ref}}$ is integrated over an interval from ${t_1 = \SI{0.01}{ms}}$ to ${t_2 = \SI{0.03}{ms}}$, which is highlighted in the top right panel of Fig.~\ref{fig:angled}.
To normalize the interface error norm~$\contourErrRel$, the integral in~\eqref{eq:contour_error} is divided by the time interval length ${(t_2 - t_1)}$ and a reference area extending over the domain width~$2a$ and an estimated characteristic height ${b = \SI{50}{\mu m}}$.
The centre right panel of Fig.~\ref{fig:angled} shows the interface error norm~$\contourErrRel$ over the minimum finite element side length~$h$.
Both the HSDI approach variants and the comparative DI model achieve a convergence order of~$O(h^1)$, which is limited by the DI formulation of the two-phase flow model in all variants.
However, for the same mesh, the HSDI approach variants are roughly one order of magnitude more accurate than the comparative model.
More precisely, to achieve the benchmark tolerance level of 1\%, an interpolated minimum finite element side length of ${h \approx \SI{1.8}{\mu m}}$ is required for the HSDI approach compared to ${h \approx \SI{0.19}{\mu m}}$ of the comparative DI model.
Thus, the purely DI model requires a mesh resolution that is at least 9 times finer to achieve the same level of accuracy.

To illustrate the deviation in the metal--gas interface contour~$\Bx_\Gamma$ for the benchmark accuracy level, the bottom panel of Fig.~\ref{fig:angled} shows the interface contours at ${t = \SI{0.03}{ms}}$ for the two aforementioned spatial resolutions.
Here, the large deviation of the DI model with ${h = \SI{1.56}{\mu m}}$ becomes visible, while the HSDI approach is very close to the reference solution for both considered mesh sizes.

\subsubsection{Unstable vapor depression}
\label{sec:angled_unstable}

This example aims to investigate the unstable keyhole dynamics that develop after the initial formation of a stable vapor depression.
We adapt the previous setup from Section~\ref{sec:angled_stable} by increasing the laser power to ${\laserPower = \SI{200}{W}}$ and the length parameter to ${a = \SI{200}{\mu m}}$, enlarging the domain.
The unstable keyhole dynamics necessitate a smaller time step size at high spatial resolutions for stability, so we employ a constant time step size of ${\dt = \SI{2.5e-9}{s}}$ for all simulations in this section.

\begin{figure}[tb!]
	\centering
	\includegraphics{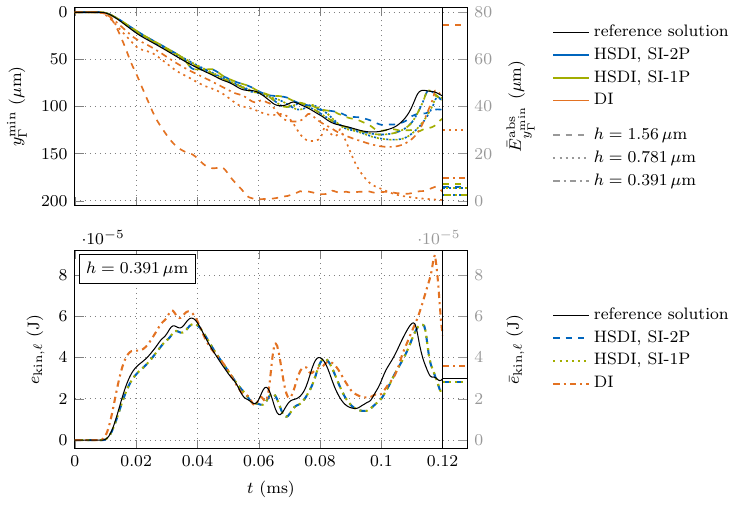}
	\caption{
		2D unstable vapor depression:
		comparison between the HSDI approach with \mbox{SI-2P}, \mbox{SI-1P}, and the purely DI melt pool model~\cite{much2024improved};
		(top)~vapor depression depth~$\interfaceDepthTwoD$ over time for different mesh resolutions and time-averaged deviation~$\interfaceDepthTwoDAvgAbs$ \eqref{eq:ymin_avg_abs};
		(bottom)~kinetic energy of the metal phase~$\kinEnergyDensityL$ \eqref{eq:metal_kin_energy} over time and time-averaged kinetic energy~$\kinEnergyDensityLAvg$ \eqref{eq:metal_kin_energy_avg} for the finite element side length ${h = \SI{0.391}{\mu m}}$.
	}
	\label{fig:unstable}
\end{figure}

The top panel of Fig.~\ref{fig:unstable} shows the depth of the laser-induced vapor depression~$\interfaceDepthTwoD$ over time.
Within the shown time interval, the reference solution shows a fairly steady vapor depression growth with minor fluctuations until ${t \approx \SI{0.1}{ms}}$, followed by a subsequent major bounce back, which initiates the high-frequency oscillations typical of melt pool dynamics.
The HSDI approach variants follow the same behaviour, and an increased spatial resolution results in a better correlation with the reference solution.
For the purely DI melt pool model, only the result obtained with the finest spatial resolution, i.e. ${h = \SI{0.391}{\mu m}}$, resembles the behaviour of the reference solution.
The right axis of the graph shows the time-averaged deviation~$\interfaceDepthTwoDAvgAbs$ of the vapor depression depth from the reference solution~$\interfaceDepthTwoDRef$, according to
\begin{align} \label{eq:ymin_avg_abs}
	\interfaceDepthTwoDAvgAbs = \frac{
		\int_{t_1}^{t_2} \left| \interfaceDepthTwoD - \interfaceDepthTwoDRef \right| \diffd t
	}{t_2 - t_1},
\end{align}
in the time interval from ${t_1 = 0}$ to ${t_2 = \SI{0.12}{ms}}$.
Evidently, the HSDI variants achieve a smaller deviation~$\interfaceDepthTwoDAvgAbs$ with the coarsest resolution (${h = \SI{1.56}{\mu m}}$) than the DI model with the finest resolution (${h = \SI{0.391}{\mu m}}$).

The bottom panel of Fig.~\ref{fig:unstable} shows the temporal evolution of the kinetic energy in the metal phase
\begin{align} \label{eq:metal_kin_energy}
	\kinEnergyDensityL = \frac{1}{2} \bar{a} \int_{\OmegaL} \rhoEff \vel^2 \diffd \Bx
\end{align}
for the finest spatial resolution.
In~\eqref{eq:metal_kin_energy}, we assume a reference depth of ${\bar{a} = \SI{1}{m}}$ for the 2D problem.
The system exhibits non-repeating, chaotic oscillations.
While the results of the HSDI approach variants resemble the oscillation behaviour of the reference solution, the purely DI model significantly deviates, starting at ${t \approx \SI{0.06}{ms}}$.
In the bottom panel of Fig.~\ref{fig:unstable}, the right axis of the graph shows the time-averaged kinetic energy in the metal phase~$\kinEnergyDensityLAvg$, according to
\begin{align} \label{eq:metal_kin_energy_avg}
	\kinEnergyDensityLAvg = \frac{
		\int_{t_1}^{t_2} \kinEnergyDensityL \diffd t,
	}{t_2 - t_1},
\end{align}
over the time interval from ${t_1 = 0}$ to ${t_2 = \SI{0.12}{ms}}$.
This time-averaged measure captures the chaotic system behaviour in a representative scalar number.
Here, the HSDI variants exhibit an error of 5\%, while the DI model results in an error of approx. 20\%.
Thus, the novel HSDI variant yields an improved accuracy by a factor of approx. four.
\\

At any given spatial discretization, the two-phase flow computation is identical for the HSDI approach variants and the purely DI melt pool model.
However, different formulations are employed by the HSDI and DI models to calculate the interface temperature as required for the evaluation of the evaporation-induced recoil pressure force~$\fRecoilPressureDiffuse$.
The substantial improvement in accuracy observed for the HSDI approaches can therefore be attributed exclusively to a more reliable prediction of the temperature field, particularly the interface temperature and the resulting diffuse-interface forces.
In this sense, an accurate interface representation enables the SI heat transfer model variants to supply the DI two-phase flow model with a markedly more accurate interface temperature than a purely DI heat transfer model, thereby allowing for a significantly more accurate description of melt pool dynamics.
In practically relevant simulations, the improved accuracy enables the prediction of melt pool dynamics with significantly higher certainty at reasonable spatial resolutions.

\subsection{3D stationary laser-induced heating of a bare metal plate}
\label{sec:melt_pool_3D}

In this section, we demonstrate the robustness of the HSDI approach and the capabilities of the numerical framework by means of a proof-of-principle simulation based on a practically relevant scenario of laser--metal interaction.
To that, we consider the stationary laser-induced heating of a bare metal plate and reproduce the corresponding experimental setup by Cunningham et al.~\cite{cunningham2019keyhole}, which was also replicated, e.g., in \mbox{\cite{zhu2021mixed, much2024improved, schreter2025consistent, meier2021novel}}.
Here, a \tiSixFour metal plate is illuminated by a vertical, stationary laser, leading to an unstable keyhole.
We consider the 3D domain ${\Omega = \{\Bx \in [-a,a]^3\}}$, with ${a = \SI{200}{\mu m}}$, and an initial signed distance ${\distance(\Bx) = -z}$, placing the initial metal--gas interface~$\Gamma$ on the $xy$-plane.
In the initial state, the temperature is uniform at ${T_0 = \SI{500}{K}}$, and the velocity is at rest ${\vel_0 = \Bzero}$.
At the top (${z = a}$) and bottom(${z = -a}$) boundaries, the temperature is prescribed to ${\hat{T} = \SI{500}{K}}$ while the remaining vertical boundaries are considered adiabatic.
Slip conditions~\eqref{eq:slip_BC} are assumed on all boundaries.
The metal--gas interface~$\Gamma$ is subject to a laser heat flux~$\qLaser$ according to~\eqref{eq:laser_gauss_sharp} with the laser power ${\laserPower = \SI{156}{W}}$, radius ${\laserRadius = \SI{70}{\mu m}}$, position ${\laserPosition = \Bzero}$, and direction~$\laserDirection = -\Be_z$ corresponding to the negative $z$-direction.
We employ material parameters representing \tiSixFour, as listed in Table~\ref{tab:param_ti64}.

For this example, we consider the \mbox{SI-1P} heat transfer model, cf. Section~\ref{sec:one_phase_description}, to model the heat transfer in the metal phase, neglecting the ambient gas.
We use an interface thickness of ${\interfaceThickness = \SI{15}{\mu m}}$ for the DI flow model.
Moreover, we employ adaptive mesh refinement to ensure a high spatial resolution of the interface region and the melt phase while minimizing the overall number of finite elements.
The base mesh, consisting of uniform hexahedral finite elements with a side length of ${h_{\*{max}} = \SI{12.5}{\mu m}}$, is locally refined to ${h_{\*{min}} = \SI{1.56}{\mu m}}$, resulting in ${\numberOfElementsInInterface \approx 9.6}$ finite elements across the interface.
Extrapolating from the \emph{1D laser-induced heating of a static surface} benchmark in Section~\ref{sec:onedim_benchmark}, this resolution should allow to predict the evaporation-induced recoil pressure within a tolerance of 0.2\% (see bottom right panel of Fig.~\ref{fig:onedim}).
In addition, as shown for the \emph{2D laser-induced stable vapor depression} in Section~\ref{sec:angled_stable}, it should suffice to resolve the melt pool surface geometry with an error below 1\%.
A time step size of ${\dt = \SI{2e-8}{s}}$ is used for time integration.

\begin{figure}[tb!]
	\centering
	\includegraphics{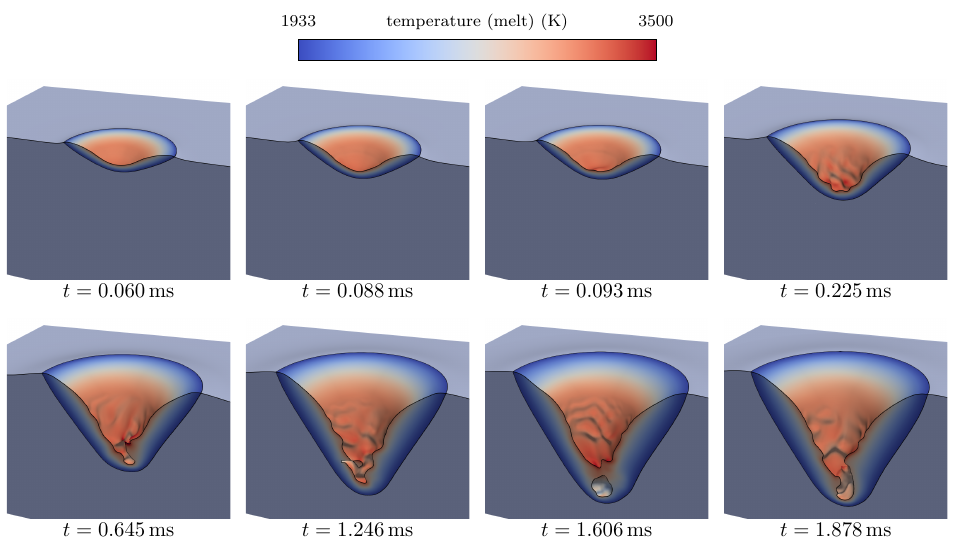}
	\caption{
		Stationary laser-induced heating of a bare metal plate:
		time series illustrating a sectional view of the melt pool shape with the temperature field of the melt.
	}
	\label{fig:melt_pool_3D_frames}
\end{figure}

\begin{figure}[tb!]
	\centering
	\includegraphics{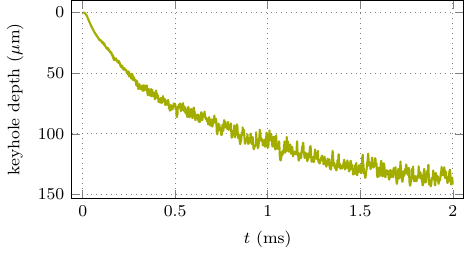}
	\caption{
		Stationary laser-induced heating of a bare metal plate:
		maximum keyhole depth over time.
	}
	\label{fig:keyhole_depth}
\end{figure}

Fig.~\ref{fig:melt_pool_3D_frames} shows a time-series of a sectional view in the melt pool, depicting the metal--gas interface~$\Gamma$ and the temperature of the melt phase.
Additional illustrations are shown in \ref{sec:melt_pool_3D_illustrations}.
The simulation reproduces typical melt pool characteristics observed in the experiment by Cunningham et al.~\cite{cunningham2019keyhole}:
Initially, a stable vapor depression forms due to the increasing recoil pressure.
As the vapor depression grows, it becomes unstable at approx. ${t = \SI{0.09}{ms}}$ in the simulation.
The interplay between the laser absorption, evaporation-induced recoil pressure force, and temperature-dependent surface tension results in high-frequency oscillations that amplify with time.
In the highly dynamic two-phase flow at the bottom of the melt pool, numerous breakup and coalescence events occur, demonstrating that the HSDI approach is capable of robustly handling these phenomena.
With time, the vapor depression becomes deeper, forming a keyhole.
Fig.~\ref{fig:keyhole_depth} shows the keyhole depth over time.
We deliberately refrain from comparing this result with the experiment, as we would not expect an agreement due to the simplified laser absorption model.
The laser heat input is the catalyst of the melt pool dynamics and must therefore precisely replicate physical mechanisms, such as multiple reflections via a ray-tracing scheme and temperature-dependent absorptivity of metals \mbox{\cite{minissale2018effect, simonds2021causal}}.
Furthermore, we hypothesize that the temperature-dependent surface tension and the resulting Marangoni convection have a significant impact on the molten metal surface geometry during the stable vapor depression phase.
However, precise measurement of temperature-dependent parameters remains challenging for typical laser-based metal processing conditions, and the selected values are subject to uncertainty, potentially influencing the onset of the vapor depression.
These aspects are not the focus of the present work, which primarily deals with the numerical formulation of an HSDI approach and its accuracy.
The extension of this model to include a ray-tracing scheme and other physical effects, such as wetting, will be addressed in our future research.
Furthermore, we intentionally refrain from comparing with a standard purely DI melt pool model, because that approach demands prohibitively high computational effort to achieve spatial convergence, as we discussed in \cite{much2024improved}.

In the following, we estimate the Reynolds number $\Re$ and the element Péclet number $\Pe$ for the present simulation.
\begin{align}
	\Re = \frac{\rhoL\,u\,L}{\inLiquid{\mu}}
	,\quad
	\Pe = \frac{u\,h_{\*{min}}\,\rhoL}{2\inLiquid{\mu}}
\end{align}
Here, $u$ denotes a characteristic velocity.
For the characteristic length~$L$, we choose the laser beam diameter, i.e., ${L = 2\laserRadius}$, since the melt pool dynamics are mostly confined to the laser spot size.
Moreover, we consider the minimum element side length~$h_{\*{min}}$, since our adaptive mesh refinement strategy ensures the highest resolution in the melt phase.
The relevant flow velocity in the predominant region of the melt pool is roughly in the order of ${u = \SI{1}{m\per s}}$.
Using this value as an estimate, the Reynolds number results in ${\Re = 163}$, which is moderate and supports our model assumptions.
The element Péclet number is ${\Pe = 0.911}$, meaning advection is not dominating, and advection stabilization is not yet required, cf.~\cite{kronbichler2018fast}.
However, the chaotic nature of the melt pool dynamics generates occasional high velocity peaks that can reach single maximal values of ${u_{\*{max}} = \SI{15}{m\per s}}$ and are typically located at the metal--gas interface.
For these peak velocities, the Reynolds number is estimated by ${\Re = 2452}$, which falls within the transitional flow regime.
Here, the element Péclet number ${\Pe = 13.7}$ suggests that advection is dominating.
In this proof-of-principle simulation, we did not employ advection stabilization.
For future research, especially for simulations with increased dynamics, advection stabilization methods, such as SUPG, are recommended.

Using the proof-of-principle example of 3D stationary laser-induced heating of a bare metal plate, we demonstrated the capabilities and robustness of the numerical framework with the HSDI approach.
The SI modelling approach for heat transfer enables the prediction of interface temperature with high accuracy, which is crucial for an accurate computation of temperature-dependent interface fluxes, such as surface tension~\eqref{eq:surface_tension_diffuse} and the evaporation-induced recoil pressure~\eqref{eq:recoil_pressure}.
The latter is especially sensitive to the interface temperature due to its exponential dependency, and represents the main driving force behind melt pool dynamics and keyhole formation.
With the DI flow model, the melt pool's complex and highly dynamic two-phase flow, including breakup and coalescence phenomena, is robustly and accurately represented.

\section{Conclusion}
\label{sec:conclusion}

Accurate prediction of melt pool dynamics in laser-based processing of metals, such as laser beam welding or additive manufacturing via laser powder bed fusion (PBF-LB/M), demands highly precise determination of the interface temperature, as the dominant driving forces, such as the evaporation-induced recoil pressure, are highly sensitive to the melt pool surface temperature.
Standard diffuse-interface (DI) formulations, though robust, introduce significant temperature errors at practically feasible spatial resolutions, as critically discussed in our previous work~\cite{much2024improved}.
As a result, model calibration procedures often compensate numerical rather than physical deficiencies.

To overcome these limitations, this work proposed a hybrid sharp--diffuse interface (HSDI) approach that couples, for the first time, a sharp-interface (SI) CutFEM heat transfer formulation with a DI two-phase flow model.
This combination is particularly well-suited for high-fidelity modelling of laser--metal interactions:
The SI formulation provides a high thermal accuracy at the melt pool surface, whereas the DI formulation offers robust and efficient handling of the complex two-phase flow, including complex breakup and coalescence scenarios.
Based on operator splitting and by extending the interface temperature within a narrow band around the (sharp) interface and embedding it into regularized force terms, our approach achieves a seamless and robust coupling between the SI thermal and DI flow submodels.

We validated the HSDI approach using comprehensive test cases and demonstrated substantial accuracy gains compared to purely DI models.
In purely thermal test cases representative of laser--metal interactions, the proposed SI thermal model consistently attains second-order convergence and, critically, at least two orders of magnitude higher accuracy than a standard DI model based on the same spatial discretization.
To achieve a 1\% error in interface temperature and evaporation-induced recoil pressure, the SI approach allowed up to two orders of magnitude coarser mesh resolutions than the DI approach.
However, the accuracy of the coupled thermo-hydrodynamics may be constrained by the two-phase flow formulation for highly dynamic 3D scenarios.
A novel benchmark example has been proposed mimicking the coupled thermal two-phase flow characteristics in laser-based metal processing.
For this more complex, coupled thermo-hydrodynamics problem, the novel HSDI approach yields an overall accuracy gain of roughly one order of magnitude in the predicted vapor depression geometry as compared to the standard DI model.
We further analyzed the effect of restricting the SI thermal problem to the metal phase, neglecting heat transfer in the gas phase.
We found that this approach is well-justified for unresolved evaporation models.
It preserves accuracy compared to the fully coupled two-phase approach, thereby enabling more efficient modelling.
Finally, we demonstrated the capabilities, robustness, and practical applicability of the proposed numerical framework using a 3D proof-of-principle simulation example of stationary laser-induced melting.
The main findings of this work are summarized as follows:
\begin{itemize}
	\item This work proposed a novel HSDI approach for thermal two-phase flow modelling that is tailored to the stringent requirements of laser-induced melt pool dynamics with rapid evaporation.
	\item In practically relevant 2D benchmarks, the SI thermal model exhibited second-order spatial convergence, enabling finite element sizes two orders of magnitude larger than standard DI approaches for 1\% accuracy.
	\item The HSDI approach is facilitated by a robust, seamless coupling between the SI thermal and DI two-phase flow submodels by extending the interface temperature across the diffuse transition region.
	\item In a novel benchmark example representative of laser--metal interactions, the HSDI approach achieved an accuracy one order of magnitude higher than a standard DI model on the same mesh.
\end{itemize}

Future work will incorporate more detailed physics of laser--material interaction, including temperature-dependent absorptivity and multiple reflections at the sharp metal--gas interface.
The methodology is directly applicable to processes involving strongly localized heat sources, such as laser and electron beam powder bed fusion, directed energy deposition, laser beam welding, cutting, and more general problems of temperature-driven flow.

\section*{Funding sources}

Financial support for this research was provided by the European Research Council through the ERC Starting Grant ExcelAM under award number 101117579.

\section*{CRediT authorship contribution statement}

\textbf{Nils Much:}
Writing -- original draft,
Writing -- review \& editing,
Conceptualization,
Methodology,
Software,
Investigation,
Formal analysis,
Data curation,
Visualization.
\textbf{Andreas Koch:}
Writing -- review \& editing,
Software,
Data curation,
Visualization.
\textbf{Christoph Meier:}
Writing -- review \& editing,
Conceptualization,
Methodology,
Supervision,
Project administration,
Resources,
Funding acquisition.
\textbf{Magdalena Schreter-Fleischhacker:}
Writing -- review \& editing,
Conceptualization,
Methodology,
Software,
Data curation,
Visualization,
Supervision,
Project administration,
Funding acquisition.

\section*{Declaration of competing interest}

The authors declare that they have no known competing financial interests or personal relationships that could have appeared to influence the work reported in this paper.

\section*{Acknowledgments}

The authors acknowledge collaboration with Maximilian Bergbauer, Hélène Papillon-Laroche, Bruno Blais, Florian Kummer, Peter Munch and Martin Kron\-bich\-ler, as well as the \texttt{deal.II} community.

\appendix

\section{Face-based ghost-penalty stabilization}
\label{sec:ghost_penalty}

\begin{figure}[H]
	\centering
	\subfloat[Ghost-penalty \hfill \: faces~$\inMetal{\mathcal{F}}$ for domain $\OmegaL$.]{
		\includegraphics[height=0.21\linewidth]{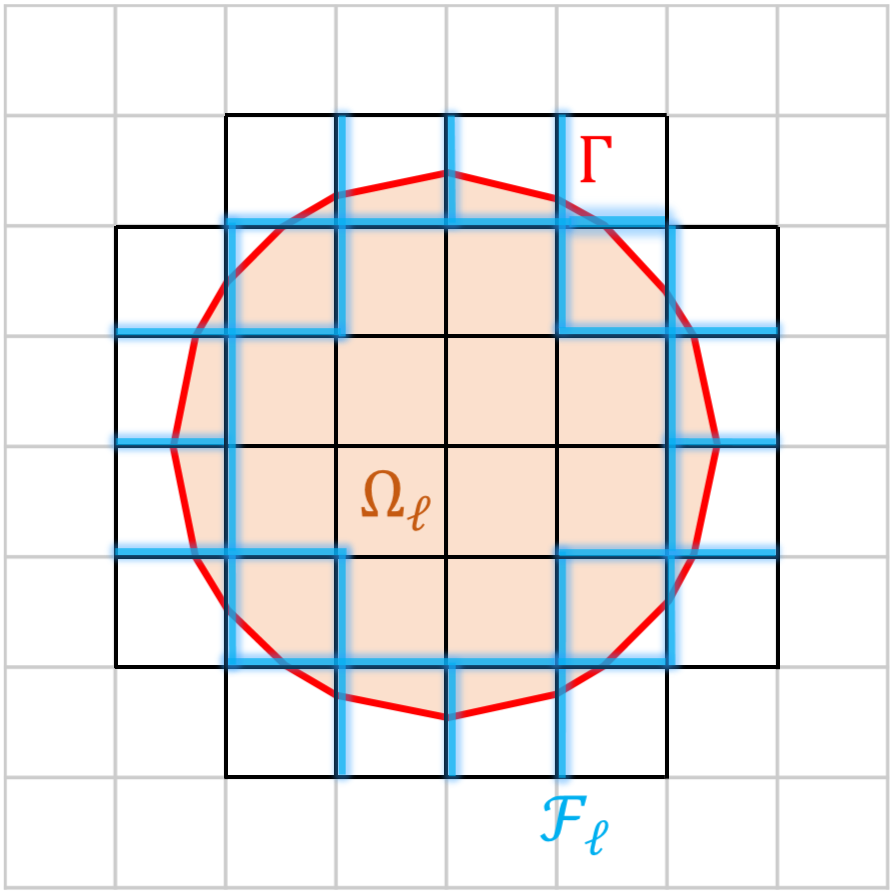}}
	\hspace{3mm}
	\subfloat[Ghost-penalty \hfill \: faces~$\inGas{\mathcal{F}}$ for domain $\OmegaG$.]{
		\includegraphics[height=0.21\linewidth]{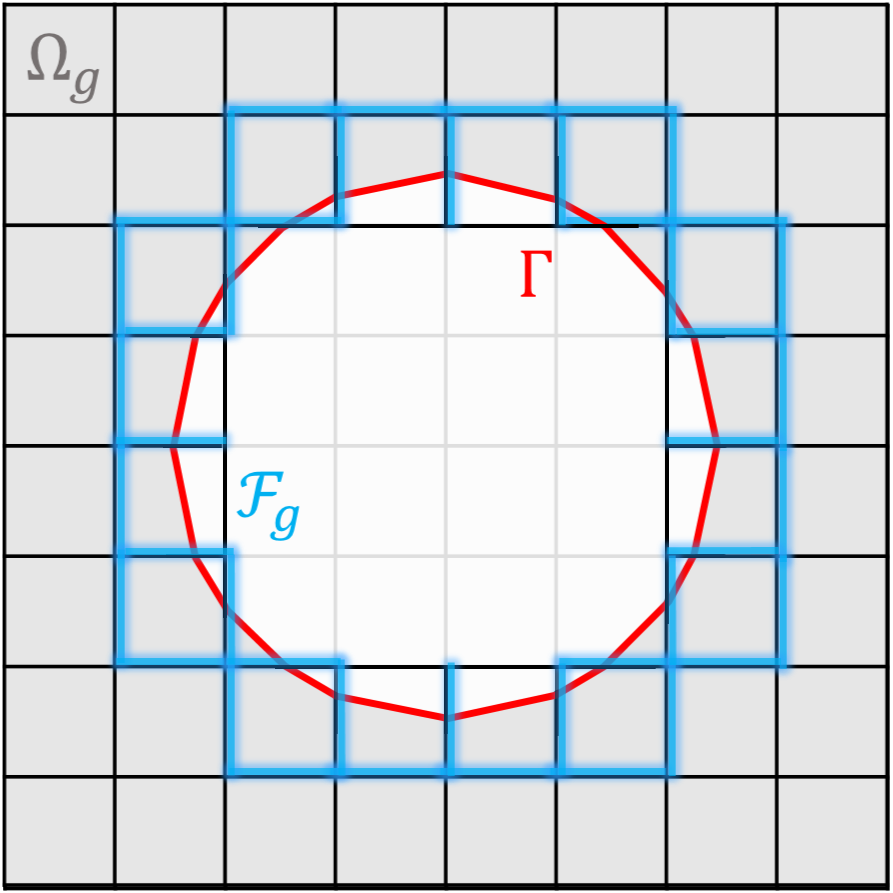}}
	\caption{
		Visualization of the set of ghost-penalty faces $\inMetal{\mathcal{F}}$ for subdomain $\OmegaL$ in (a), and $\inGas{\mathcal{F}}$ for subdomain $\OmegaG$ in (b), respectively.
		The ghost-penalty faces are highlighted in blue.}
	\label{fig:ghost_penalty_faces}
\end{figure}

In intersected finite elements, domain integrals in~$\OmegaL$ and~$\OmegaG$ must only consider the part of the element that lies inside the respective domain.
Since the metal--gas interface~$\Gamma$ may cut finite elements in any location, the intersections can get arbitrarily small, leading to very bad condition numbers of the mass and stiffness matrices.
To mitigate this issue, we apply face-based ghost-penalty stabilization~\cite{burman2010ghost} to the mass and stiffness.
The stabilization functions~$\GPstabilFunctionL$ and~$\GPstabilFunctionG$ are defined as
\begin{align} \label{eq:stabilization_function_j}
	\GPstabilFunctionS(\Ts,\vS)
	= \sum_{\face \in \setOfFacesS} \sum_{\GPPIndex=1}^p \frac{h^{2\GPPIndex+1}}{(2\GPPIndex+1)(\GPPIndex!)^2}
	\biggl( \left[ \partial_{\Bn_\face}^\GPPIndex \Ts \right]_\face \, , \left[ \partial_{\Bn_\face}^\GPPIndex \vS \right]_\face \biggr)_\face
	\quad \text{for } \phaseIndex \in \metalAndGas,
\end{align}
where~$\face$ is the domain of a finite element face that is in one of the sets of element faces~$\inMetal{\setOfFaces}$ and~$\inGas{\setOfFaces}$, which contain all faces of intersected finite elements except element faces that do not cover parts of the respective domain~$\OmegaL$ or~$\OmegaG$~\cite{ludvigsson2018high} as shown in Fig.~\ref{fig:ghost_penalty_faces}.
Furthermore, $p$ is the degree of the Lagrange polynomial shape functions, $\Bn_\face$ is the outward-pointing unit normal vector of a face~$\face$, $\partial_{\Bn_\face}^\GPPIndex$ is the $\GPPIndex$-th order normal derivative on a face~$\face$, and the operator ${\left[ x \right]_\face = \left. x \right|_{\face^+} - \left. x \right|_{\face^-}}$ denotes the jump over a face~$\face$.

\section{Solution projection between function spaces for moving interfaces}
\label{sec:ghost_penalty_extrapol}

\begin{figure}[tb!]
	\centering
	\subfloat[Active submesh~$\TauL^\timeStepIndex$ at~time step $t^\timeStepIndex$.]{
		\includegraphics[height=0.21\linewidth]{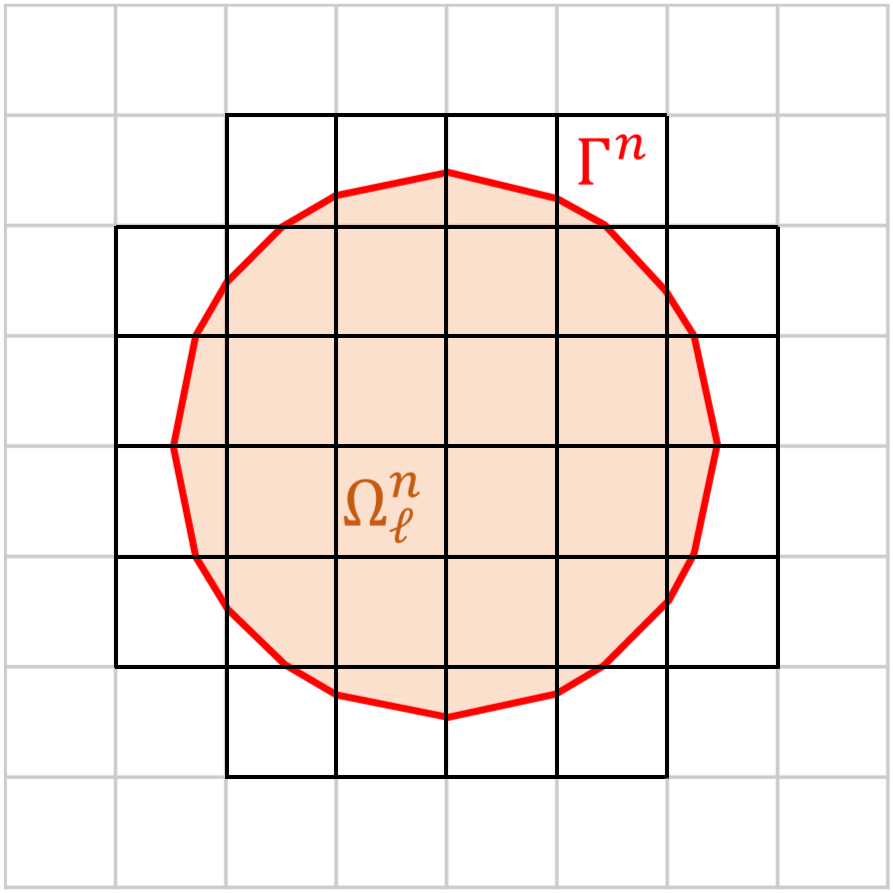}}
	\hspace{3mm}
	\subfloat[Active submesh~$\TauL^{\timeStepIndex+1}$ at~time step $t^{\timeStepIndex+1}$.]{
		\includegraphics[height=0.21\linewidth]{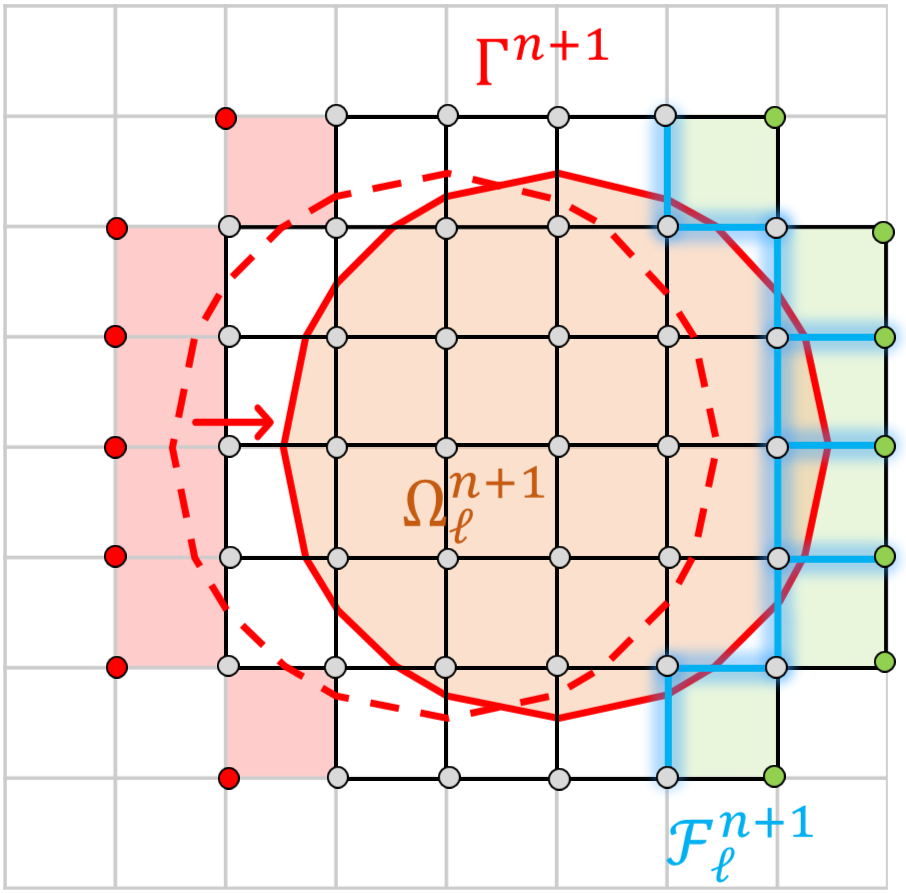}}
	\hspace{3mm}
	\subfloat[3D wrap illustration of the 2D \hfill \: extrapolation example.]{
		\includegraphics[height=0.21\linewidth]{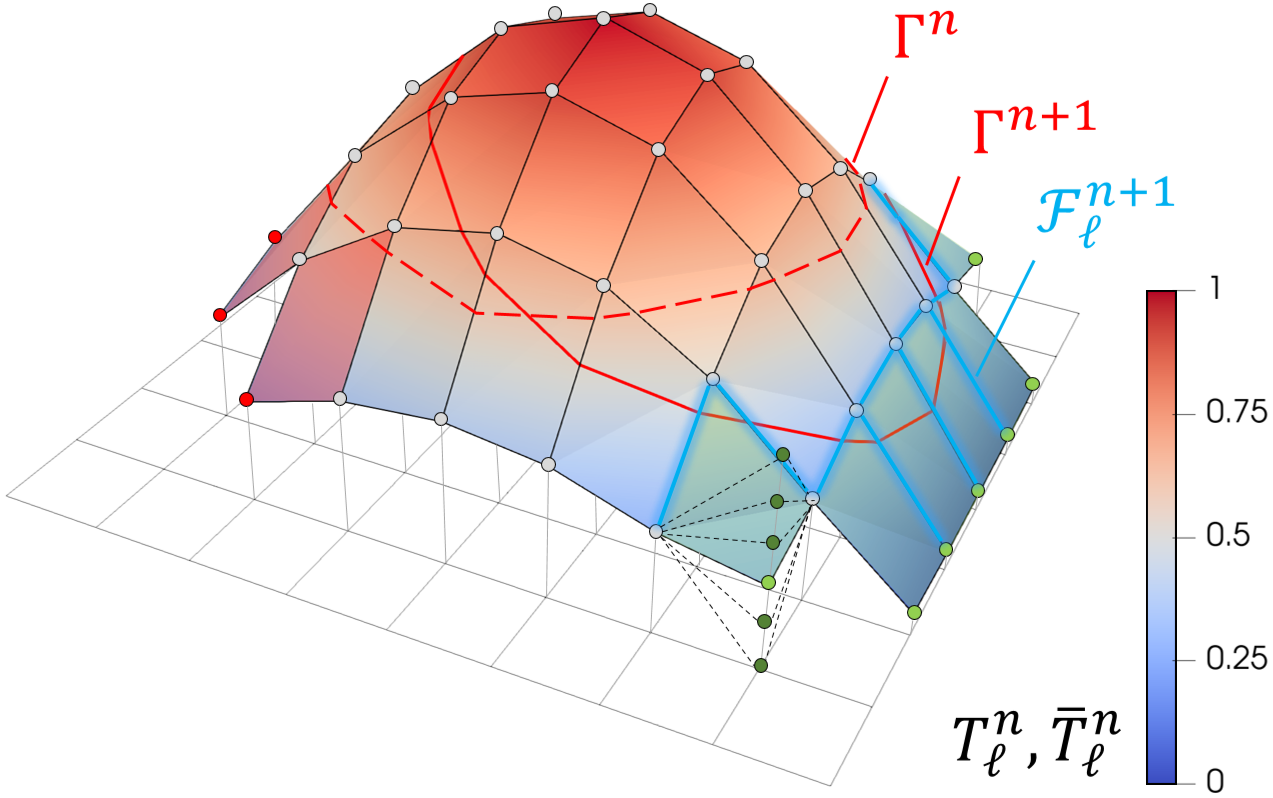}}
	\caption{
		Sketches for visualizing the solution extrapolation process between the differing active submesh topologies ${\TauL^\timeStepIndex \neq \TauL^{\timeStepIndex+1}}$ in two subsequent time steps~$t^\timeStepIndex$ and~$t^{\timeStepIndex+1}$.
		For the illustration, we only consider the subdomain~$\OmegaL$ in our example.
		In (a), the original state at~$t^\timeStepIndex$ is shown and in (b), the horizontally moved interface~$\Gamma^{\timeStepIndex+1}$ with the new active submesh~$\TauL^{\timeStepIndex+1}$ is shown.
		The grey points represent the solution, which can be adopted to the new projected solution ${\inLiquid{\bar{T}}^\timeStepIndex = \Tl^\timeStepIndex}$ via constraints (scenario~1).
		The red-marked solution is not relevant anymore, as it does not belong to the current active submesh (scenario~2).
		On the contrary, the green marked solution is new and needs an extrapolation process, so an accurate solution value~$\inLiquid{\bar{T}}^\timeStepIndex$ can be predicted.
		The relevant set of ghost-faces for the solution projection process~$\setOfFacesL^{\timeStepIndex+1}$ is highlighted in blue.
		In (c), a 3D warp illustration is depicted, showing the solution fields~$\Tl^\timeStepIndex$ and~$\inLiquid{\bar{T}}^\timeStepIndex$ for a 2D sine-bump-shaped solution.
	}
	\label{fig:Sketch ghost-penalty extrapolation 2D}
\end{figure}

Consider a structured triangulation~$\triangulation$ that covers the domain~$\Omega$.
We distinguish between the metal triangulation~$\TauL \in \triangulation$ that includes all finite elements that have some part in $\OmegaL$ and the ambient gas triangulation~$\TauG \in \triangulation$ that includes all finite elements that have some part in~$\OmegaG$.
The set of interface elements ${\TauGamma = \TauL \cap \TauG}$ includes all finite elements intersected by the interface~$\Gamma$.
In the following, the superscript~$\timeStepIndex$ indicates the time step number and the subscript ${\phaseIndex \in \metalAndGas}$ indicates the phase.
When dealing with a moving interface~$\Gamma$, we encounter a problem if the solution at the old time step $T_\phaseIndex^\timeStepIndex$ and the solution at the new time step $T_\phaseIndex^{\timeStepIndex+1}$ differ in their solution function spaces.
This occurs if the active subdomain meshes change their topology, i.e., ${\TauS^{\timeStepIndex+1} \neq \TauS^\timeStepIndex}$, between two subsequent time steps~$t^\timeStepIndex$ and~$t^{\timeStepIndex+1}$.
To address this issue, we apply the so-called ghost-penalty extrapolation procedure proposed by Schott~\cite{schott2017stabilized}.
We aim for a \emph{virtual} solution~$\bar{T}_\phaseIndex^\timeStepIndex$, representing the solution~$T_\phaseIndex^\timeStepIndex$ on the new active submesh topology~$\TauS^{\timeStepIndex+1}$ according to the moved interface position~$\Gamma^{\timeStepIndex+1}$ at time step $t^{n+1}$.
For a moving interface ${\Gamma^\timeStepIndex \rightarrow \Gamma^{\timeStepIndex+1}}$, we can distinguish three scenarios for the finite elements ${\cell_\phaseIndex \in \TauS}$ that were physically relevant for the subdomain~$\Omega_\phaseIndex^\timeStepIndex$ at the time step~$t^\timeStepIndex$.
\\
\\
\noindent
\textit{Scenario 1: Interface motion within a cut finite element or remaining a non-intersected physically relevant finite element with equal solution function space}
\\\indent
For interface motions within the same finite element~$\cell_\phaseIndex$, i.e., ${\cell_\phaseIndex\in\TauGamma^\timeStepIndex \cap \TauGamma^{\timeStepIndex+1}}$, both finite element function spaces of the two subsequent time steps are based on the same function basis, so no extrapolation or any further special treatment is required for this scenario.
The solution can easily be adopted ${\bar{T}_\phaseIndex^\timeStepIndex = T_\phaseIndex^\timeStepIndex}$.
The same holds for finite elements, which are not intersected at both subsequent time steps, and belong to the physically relevant subdomain, i.e. ${\cell_\phaseIndex\in (\TauS^\timeStepIndex \setminus \TauGamma^\timeStepIndex) \cap (\TauS^{\timeStepIndex+1} \setminus \TauGamma^{\timeStepIndex+1})}$.
\\
\\
\textit{Scenario 2: Interface motion across nodes with reduced solution function space}
\\\indent
Suppose the interface moves across a finite element boundary~$\partial\cell_\phaseIndex$, so that the corresponding finite element ansatz functions are not required anymore for representing the discrete projected solution~$\bar{T}_\phaseIndex^\timeStepIndex$, i.e., ${\cell_\phaseIndex\in\TauGamma^\timeStepIndex \setminus \TauS^{\timeStepIndex+1}}$.
In that case, the solution space is reduced accordingly.
The topology of the active submeshes changes in this scenario ${\TauS^{\timeStepIndex+1}\neq\TauS^\timeStepIndex}$, and the solution, which is no longer required, is removed.
\\
\\
\textit{Scenario 3: Interface motion across nodes with increased solution function space}
\\\indent
If the interface moves across a finite element boundary~$\partial \cell_\phaseIndex$, we encounter the most challenging scenario.
Finite elements become physically relevant for the considered subdomain~$\OmegaS$, i.e. ${\cell_\phaseIndex\in\TauGamma^{\timeStepIndex+1} \setminus \TauS^\timeStepIndex}$, expanding the solution function space.
Since the solution of the newly activated finite elements, which were located outside the physically active domain at~$t^\timeStepIndex$, are unknown, we must determine appropriate virtual solution values without compromising the accuracy of our numerical scheme.
We apply a ghost-penalty extrapolation procedure to estimate the virtual solution values~$\bar{T}_\phaseIndex^\timeStepIndex$, aiming to minimize the normal derivative jumps on the faces between the new physically relevant finite element~$\cell_\phaseIndex$ and the neighbour elements belonging to the same considered subdomain~$\OmegaS$.
The algorithmic procedure consists of the following two steps:
\begin{itemize}
	\item Move interface ${\Gamma^\timeStepIndex\rightarrow\Gamma^{\timeStepIndex+1}}$ and determine the active subdomain meshes~$\TauS^{\timeStepIndex+1}$ at the new time step~$t^{\timeStepIndex+1}$.
	The physically irrelevant solution (scenario~2) is removed, new solution space (scenario~3) is created, and the solution from scenario~1 is adopted.
	\item Apply the ghost-penalty extrapolation procedure, similar to the face-based ghost-penalty terms in \eqref{eq:weak_heat_equation_two_phase}, for the newly arising unknown solution (scenario~3), by creating and solving a global system of equations for ${\bar{T}^\timeStepIndex = \{\inMetal{\bar{T}}^\timeStepIndex, \inGas{\bar{T}}^\timeStepIndex\}}$, such that
	\begin{align*}
		\sum_{\phaseIndex \in \metalAndGas}
		\sum_{\face \in \setOfFacesS^{n+1}}
		\sum_{\GPPIndex=1}^p
		\frac{\condS \, h^{2\GPPIndex-1}}{(2\GPPIndex+1)(\GPPIndex!)^2}
		\biggl( \left[ \partial_{\Bn_\face}^\GPPIndex \bar{T}_\phaseIndex^\timeStepIndex \right]_\face \, , \left[ \partial_{\Bn_\face}^\GPPIndex \vS \right]_\face \biggr)_\face = 0
	\end{align*}
	holds for the continuous solution space.
	Note that the exponent in~$h^{2\GPPIndex-1}$ is chosen so that the relation is spatial dimension independent.
	For the extrapolation procedure, the function spaces at the new time step have to be used, corresponding to the new active submesh topologies~$\TauS^{\timeStepIndex+1}$.
	The solution that is not newly introduced is enforced to remain the same via inhomogeneous Dirichlet constraints for the system of equations $\bar{T}_\phaseIndex^\timeStepIndex = \Ts^\timeStepIndex$.
\end{itemize}
For illustrating the ghost-penalty extrapolation, we consider a 2D example of a subdomain~$\OmegaL$ as shown in Fig.~\ref{fig:Sketch ghost-penalty extrapolation 2D}a, which is moved as shown in Fig.~\ref{fig:Sketch ghost-penalty extrapolation 2D}b.
The two different active submeshes ${\TauL^\timeStepIndex \neq \TauL^{\timeStepIndex+1}}$ are shown in Fig.~\ref{fig:Sketch ghost-penalty extrapolation 2D}b from a top view and clarify the terminologies and scenarios 1, 2, and 3 from above, as well as a spatial representation of the corresponding solution fields~$\Tl^\timeStepIndex$ and~$\inLiquid{\bar{T}}^\timeStepIndex$.
Note that the presented procedure can be equally adopted for 3D problems and vector-valued approximations.

\section{Illustration of 3D stationary laser-induced heating of a bare metal plate}
\label{sec:melt_pool_3D_illustrations}

\begin{figure}[hp!]
	\centering
	\includegraphics{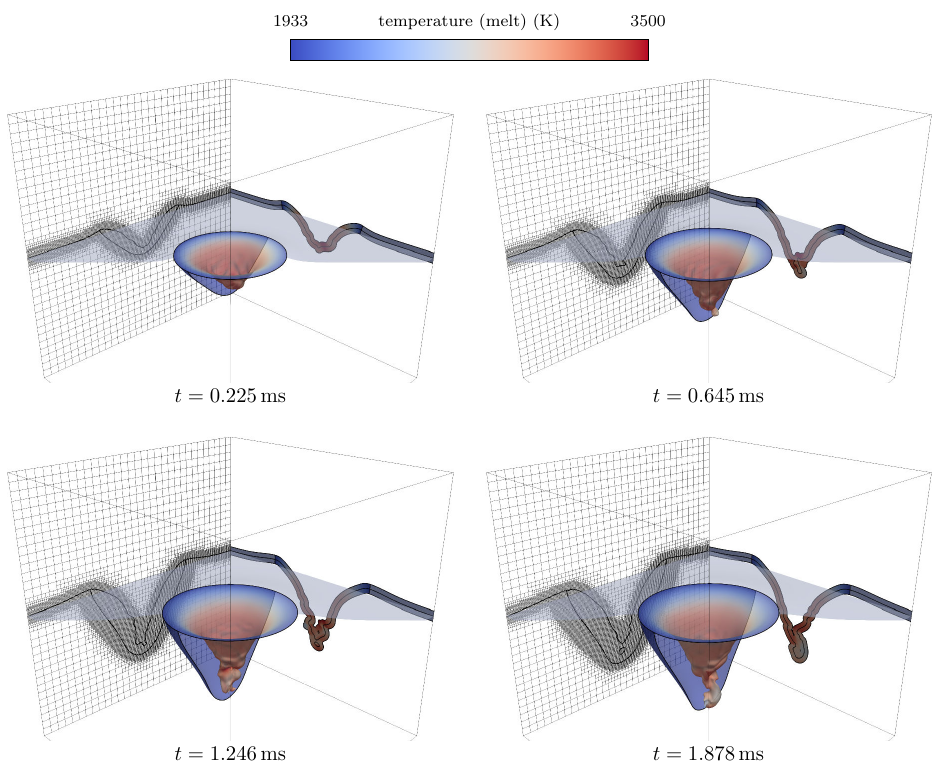}
	\caption{
		Stationary laser-induced heating of a bare metal plate:
		time series illustrating the metal--gas interface geometry, along with the solidus temperature contour.
		The adaptive mesh of the $xz$-plane is projected on the left back face, while the extended interface temperature $\TGammaExt$ at the $yz$-plane is projected on the right back face.
	}
	\label{fig:melt_pool_3D_frames_detailed}
\end{figure}

In this section, additional illustrations of the 3D stationary laser-induced heating of a bare metal plate example, cf. Section~\ref{sec:melt_pool_3D}, are shown (see Fig.~\ref{fig:melt_pool_3D_frames_detailed}).

\section{Temperature profiles of 1D laser-induced heating of a static surface}
\label{sec:onedim_profiles}

\begin{figure}[htbp!]
	\centering
	\includegraphics{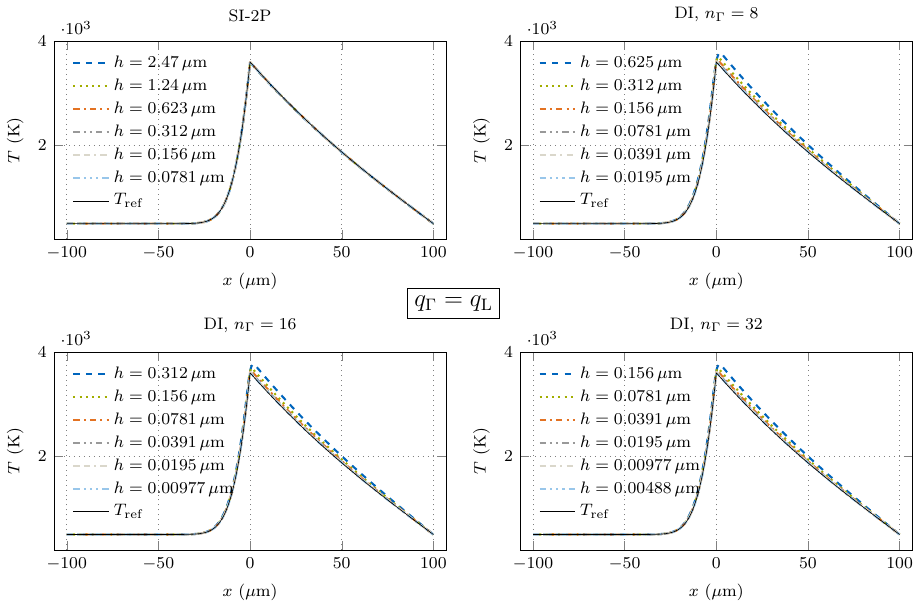}
	\caption{
		1D laser-induced heating of a static surface without evaporation-induced cooling, cf. Section~\ref{sec:onedim_no_evapor}:
		Instationary temperature profiles at $t=\SI{e-5}{s}$.
		(top left) SI-2P;
		DI heat transfer with (top right) ${\numberOfElementsInInterface = 8}$; (bottom left) ${\numberOfElementsInInterface = 16}$; (bottom right) ${\numberOfElementsInInterface = 32}$ finite elements across the diffuse transition region.
	}
	\label{fig:onedim_noevap_profiles}
\end{figure}

\begin{figure}[htbp!]
	\centering
	\includegraphics{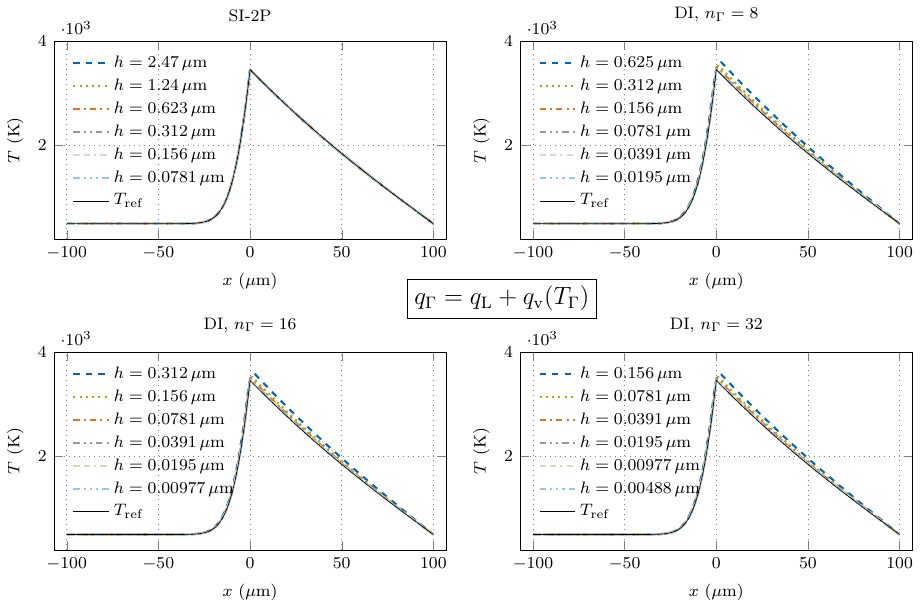}
	\caption{
		1D laser-induced heating of a static surface with evaporation-induced cooling, cf. Section~\ref{sec:onedim_with_evapor}:
		Instationary temperature profiles at $t=\SI{e-5}{s}$.
		(top left) SI-2P;
		DI heat transfer with (top right) ${\numberOfElementsInInterface = 8}$; (bottom left) ${\numberOfElementsInInterface = 16}$; (bottom right) ${\numberOfElementsInInterface = 32}$ finite elements across the diffuse transition region.
	}
	\label{fig:onedim_evap_profiles}
\end{figure}

In this section, the final temperature profiles of the 1D laser-induced heating of a static surface benchmark example, cf. Section~\ref{sec:onedim_benchmark}, are shown.
Fig.~\ref{fig:onedim_noevap_profiles} presents the temperature profiles obtained without evaporation-induced cooling, as discussed in Section~\ref{sec:onedim_no_evapor}. In contrast, Fig.~\ref{fig:onedim_evap_profiles} shows the temperature profiles incorporating evaporation-induced cooling, corresponding to the results in Section~\ref{sec:onedim_with_evapor}.

\section*{Data availability}

The code we use is openly available at \newline \href{https://github.com/MeltPoolDG/paper-2026-CMAME-hybrid-melt-pool}{https://github.com/MeltPoolDG/paper-2026-CMAME-hybrid-melt-pool}.

\FloatBarrier

\bibliography{literature.bib}

\begin{thebibliography}{10}
\expandafter\ifx\csname url\endcsname\relax
  \def\url#1{\texttt{#1}}\fi
\expandafter\ifx\csname urlprefix\endcsname\relax\def\urlprefix{URL }\fi
\expandafter\ifx\csname href\endcsname\relax
  \def\href#1#2{#2} \def\path#1{#1}\fi

\bibitem{gibson2021additive}
I.~Gibson, D.~Rosen, B.~Stucker, M.~Khorasani, {Additive Manufacturing
  Technologies}, 3rd Edition, Springer International Publishing, Cham, 2021.
\newblock \href {https://doi.org/10.1007/978-3-030-56127-7}
  {\path{doi:10.1007/978-3-030-56127-7}}.

\bibitem{meier2017thermophysical}
C.~Meier, R.~W. Penny, Y.~Zou, J.~S. Gibbs, A.~J. Hart, {Thermophysical
  phenomena in metal additive manufacturing by selective laser melting:
  fundamentals, modeling, simulation, and experimentation}, Annual Review of
  Heat Transfer 20~(1) (2017) 241--316.
\newblock \href {https://doi.org/10.1615/AnnualRevHeatTransfer.2018019042}
  {\path{doi:10.1615/AnnualRevHeatTransfer.2018019042}}.

\bibitem{li2022particle}
E.~Li, Z.~Zhou, L.~Wang, R.~Zou, A.~Yu, {Particle scale modelling of powder
  recoating and melt pool dynamics in laser powder bed fusion additive
  manufacturing: A review}, Powder Technology 409~(July) (2022) 117789.
\newblock \href {https://doi.org/10.1016/j.powtec.2022.117789}
  {\path{doi:10.1016/j.powtec.2022.117789}}.

\bibitem{cunningham2019keyhole}
R.~Cunningham, C.~Zhao, N.~Parab, C.~Kantzos, J.~Pauza, K.~Fezzaa, T.~Sun,
  A.~D. Rollett, {Keyhole threshold and morphology in laser melting revealed by
  ultrahigh-speed x-ray imaging}, Science 363~(6429) (2019) 849--852.
\newblock \href {https://doi.org/10.1126/science.aav4687}
  {\path{doi:10.1126/science.aav4687}}.

\bibitem{korner2013fundamental}
C.~K{\"{o}}rner, A.~Bauerei{\ss}, E.~Attar, {Fundamental consolidation
  mechanisms during selective beam melting of powders}, Modelling and
  Simulation in Materials Science and Engineering 21~(8) (2013) 085011.
\newblock \href {https://doi.org/10.1088/0965-0393/21/8/085011}
  {\path{doi:10.1088/0965-0393/21/8/085011}}.

\bibitem{brennan2021defects}
M.~C. Brennan, J.~S. Keist, T.~A. Palmer, {Defects in Metal Additive
  Manufacturing Processes}, Journal of Materials Engineering and Performance
  30~(7) (2021) 4808--4818.
\newblock \href {https://doi.org/10.1007/s11665-021-05919-6}
  {\path{doi:10.1007/s11665-021-05919-6}}.

\bibitem{cook2020simulation}
P.~S. Cook, A.~B. Murphy, {Simulation of melt pool behaviour during additive
  manufacturing: Underlying physics and progress}, Additive Manufacturing 31
  (2020) 100909.
\newblock \href {https://doi.org/10.1016/j.addma.2019.100909}
  {\path{doi:10.1016/j.addma.2019.100909}}.

\bibitem{hansbo2002unfitted}
A.~Hansbo, P.~Hansbo, {An unfitted finite element method, based on Nitsche's
  method, for elliptic interface problems}, Computer Methods in Applied
  Mechanics and Engineering 191~(47-48) (2002) 5537--5552.
\newblock \href {https://doi.org/10.1016/S0045-7825(02)00524-8}
  {\path{doi:10.1016/S0045-7825(02)00524-8}}.

\bibitem{chessa2002extended}
J.~Chessa, P.~Smolinski, T.~Belytschko, {The extended finite element method
  (XFEM) for solidification problems}, International Journal for Numerical
  Methods in Engineering 53~(8) (2002) 1959--1977.
\newblock \href {https://doi.org/10.1002/nme.386} {\path{doi:10.1002/nme.386}}.

\bibitem{fedkiw1999nonoscillatory}
R.~P. Fedkiw, T.~Aslam, B.~Merriman, S.~Osher, {A Non-oscillatory Eulerian
  Approach to Interfaces in Multimaterial Flows (the Ghost Fluid Method)},
  Journal of Computational Physics 152~(2) (1999) 457--492.
\newblock \href {https://doi.org/10.1006/jcph.1999.6236}
  {\path{doi:10.1006/jcph.1999.6236}}.

\bibitem{burman2010ghost}
E.~Burman, {Ghost penalty}, Comptes Rendus. Math{\'{e}}matique 348~(21-22)
  (2010) 1217--1220.
\newblock \href {https://doi.org/10.1016/j.crma.2010.10.006}
  {\path{doi:10.1016/j.crma.2010.10.006}}.

\bibitem{hirt1974arbitrary}
C.~Hirt, A.~Amsden, J.~Cook, {An arbitrary Lagrangian-Eulerian computing method
  for all flow speeds}, Journal of Computational Physics 14~(3) (1974)
  227--253.
\newblock \href {https://doi.org/10.1016/0021-9991(74)90051-5}
  {\path{doi:10.1016/0021-9991(74)90051-5}}.

\bibitem{fevrier2025simulation}
S.~F{\'{e}}vrier, E.~Fern{\'{a}}ndez, M.~Lacroix, R.~Boman, J.-P. Ponthot,
  {Simulation of melt pool dynamics including vaporization using the particle
  finite element method}, Computational Mechanics 75~(6) (2025) 1787--1815.
\newblock \href {https://doi.org/10.1007/s00466-024-02571-4}
  {\path{doi:10.1007/s00466-024-02571-4}}.

\bibitem{tryggvason2011direct}
G.~Tryggvason, R.~Scardovelli, S.~Zaleski, {Direct Numerical Simulations of
  Gas–Liquid Multiphase Flows}, Cambridge University Press, Cambridge, 2011.
\newblock \href {https://doi.org/10.1017/CBO9780511975264}
  {\path{doi:10.1017/CBO9780511975264}}.

\bibitem{pang2011three}
S.~Pang, L.~Chen, J.~Zhou, Y.~Yin, T.~Chen, {A three-dimensional sharp
  interface model for self-consistent keyhole and weld pool dynamics in deep
  penetration laser welding}, Journal of Physics D: Applied Physics 44~(2)
  (2011) 025301.
\newblock \href {https://doi.org/10.1088/0022-3727/44/2/025301}
  {\path{doi:10.1088/0022-3727/44/2/025301}}.

\bibitem{tan2013investigation}
W.~Tan, N.~S. Bailey, Y.~C. Shin, {Investigation of keyhole plume and molten
  pool based on a three-dimensional dynamic model with sharp interface
  formulation}, Journal of Physics D: Applied Physics 46~(5) (2013) 055501.
\newblock \href {https://doi.org/10.1088/0022-3727/46/5/055501}
  {\path{doi:10.1088/0022-3727/46/5/055501}}.

\bibitem{khairallah2014mesoscopic}
S.~A. Khairallah, A.~Anderson, {Mesoscopic simulation model of selective laser
  melting of stainless steel powder}, Journal of Materials Processing
  Technology 214~(11) (2014) 2627--2636.
\newblock \href {https://doi.org/10.1016/j.jmatprotec.2014.06.001}
  {\path{doi:10.1016/j.jmatprotec.2014.06.001}}.

\bibitem{khairallah2016laser}
S.~A. Khairallah, A.~T. Anderson, A.~Rubenchik, W.~E. King, {Laser powder-bed
  fusion additive manufacturing: Physics of complex melt flow and formation
  mechanisms of pores, spatter, and denudation zones}, Acta Materialia 108
  (2016) 36--45.
\newblock \href {https://doi.org/10.1016/j.actamat.2016.02.014}
  {\path{doi:10.1016/j.actamat.2016.02.014}}.

\bibitem{matthews2016denudation}
M.~J. Matthews, G.~Guss, S.~A. Khairallah, A.~M. Rubenchik, P.~J. Depond, W.~E.
  King, {Denudation of metal powder layers in laser powder bed fusion
  processes}, Acta Materialia 114 (2016) 33--42.
\newblock \href {https://doi.org/10.1016/j.actamat.2016.05.017}
  {\path{doi:10.1016/j.actamat.2016.05.017}}.

\bibitem{ly2017metal}
S.~Ly, A.~M. Rubenchik, S.~A. Khairallah, G.~Guss, M.~J. Matthews, {Metal vapor
  micro-jet controls material redistribution in laser powder bed fusion
  additive manufacturing}, Scientific Reports 7~(1) (2017) 4085.
\newblock \href {https://doi.org/10.1038/s41598-017-04237-z}
  {\path{doi:10.1038/s41598-017-04237-z}}.

\bibitem{martin2019dynamics}
A.~A. Martin, N.~P. Calta, S.~A. Khairallah, J.~Wang, P.~J. Depond, A.~Y. Fong,
  V.~Thampy, G.~M. Guss, A.~M. Kiss, K.~H. Stone, C.~J. Tassone, J.~{Nelson
  Weker}, M.~F. Toney, T.~van Buuren, M.~J. Matthews, {Dynamics of pore
  formation during laser powder bed fusion additive manufacturing}, Nature
  Communications 10~(1) (2019) 1987.
\newblock \href {https://doi.org/10.1038/s41467-019-10009-2}
  {\path{doi:10.1038/s41467-019-10009-2}}.

\bibitem{brackbill1992continuum}
J.~Brackbill, D.~Kothe, C.~Zemach, {A continuum method for modeling surface
  tension}, Journal of Computational Physics 100~(2) (1992) 335--354.
\newblock \href {https://doi.org/10.1016/0021-9991(92)90240-Y}
  {\path{doi:10.1016/0021-9991(92)90240-Y}}.

\bibitem{much2024improved}
N.~Much, M.~Schreter-Fleischhacker, P.~Munch, M.~Kronbichler, W.~A. Wall,
  C.~Meier, {Improved accuracy of continuum surface flux models for metal
  additive manufacturing melt pool simulations}, Advanced Modeling and
  Simulation in Engineering Sciences 11~(1) (2024) 16.
\newblock \href {https://doi.org/10.1186/s40323-024-00270-6}
  {\path{doi:10.1186/s40323-024-00270-6}}.

\bibitem{schreter2025consistent}
M.~Schreter-Fleischhacker, N.~Much, P.~Munch, M.~Kronbichler, W.~A. Wall,
  C.~Meier, {A consistent diffuse-interface finite element approach to rapid
  melt–vapor dynamics with application to metal additive manufacturing},
  Computer Methods in Applied Mechanics and Engineering 442 (2025) 117985.
\newblock \href {https://doi.org/10.1016/j.cma.2025.117985}
  {\path{doi:10.1016/j.cma.2025.117985}}.

\bibitem{andreotta2017finite}
R.~Andreotta, L.~Ladani, W.~Brindley, {Finite element simulation of laser
  additive melting and solidification of Inconel 718 with experimentally tested
  thermal properties}, Finite Elements in Analysis and Design 135~(July) (2017)
  36--43.
\newblock \href {https://doi.org/10.1016/j.finel.2017.07.002}
  {\path{doi:10.1016/j.finel.2017.07.002}}.

\bibitem{yan2018fully}
J.~Yan, W.~Yan, S.~Lin, G.~Wagner, {A fully coupled finite element formulation
  for liquid–solid–gas thermo-fluid flow with melting and solidification},
  Computer Methods in Applied Mechanics and Engineering 336 (2018) 444--470.
\newblock \href {https://doi.org/10.1016/j.cma.2018.03.017}
  {\path{doi:10.1016/j.cma.2018.03.017}}.

\bibitem{zhu2021mixed}
Q.~Zhu, J.~Yan, {A mixed interface-capturing/interface-tracking formulation for
  thermal multi-phase flows with emphasis on metal additive manufacturing
  processes}, Computer Methods in Applied Mechanics and Engineering 383 (2021)
  113910.
\newblock \href {https://doi.org/10.1016/j.cma.2021.113910}
  {\path{doi:10.1016/j.cma.2021.113910}}.

\bibitem{courtois2014complete}
M.~Courtois, M.~Carin, P.~{Le Masson}, S.~Gaied, M.~Balabane, {A complete model
  of keyhole and melt pool dynamics to analyze instabilities and collapse
  during laser welding}, Journal of Laser Applications 26~(4) (2014) 042001.
\newblock \href {https://doi.org/10.2351/1.4886835}
  {\path{doi:10.2351/1.4886835}}.

\bibitem{leitz2018fundamental}
K.-H. Leitz, C.~Grohs, P.~Singer, B.~Tabernig, A.~Plankensteiner, H.~Kestler,
  L.~Sigl, {Fundamental analysis of the influence of powder characteristics in
  Selective Laser Melting of molybdenum based on a multi-physical simulation
  model}, International Journal of Refractory Metals and Hard Materials
  72~(November 2017) (2018) 1--8.
\newblock \href {https://doi.org/10.1016/j.ijrmhm.2017.11.034}
  {\path{doi:10.1016/j.ijrmhm.2017.11.034}}.

\bibitem{queva2020numerical}
A.~Queva, G.~Guillemot, C.~Moriconi, C.~Metton, M.~Bellet, {Numerical study of
  the impact of vaporisation on melt pool dynamics in Laser Powder Bed Fusion -
  Application to IN718 and Ti–6Al–4V}, Additive Manufacturing 35~(February)
  (2020) 101249.
\newblock \href {https://doi.org/10.1016/j.addma.2020.101249}
  {\path{doi:10.1016/j.addma.2020.101249}}.

\bibitem{panwisawas2017mesoscale}
C.~Panwisawas, C.~Qiu, M.~J. Anderson, Y.~Sovani, R.~P. Turner, M.~M. Attallah,
  J.~W. Brooks, H.~C. Basoalto, {Mesoscale modelling of selective laser
  melting: Thermal fluid dynamics and microstructural evolution}, Computational
  Materials Science 126 (2017) 479--490.
\newblock \href {https://doi.org/10.1016/j.commatsci.2016.10.011}
  {\path{doi:10.1016/j.commatsci.2016.10.011}}.

\bibitem{bayat2019multiphysics}
M.~Bayat, S.~Mohanty, J.~H. Hattel, {Multiphysics modelling of lack-of-fusion
  voids formation and evolution in IN718 made by multi-track/multi-layer
  L-PBF}, International Journal of Heat and Mass Transfer 139 (2019) 95--114.
\newblock \href {https://doi.org/10.1016/j.ijheatmasstransfer.2019.05.003}
  {\path{doi:10.1016/j.ijheatmasstransfer.2019.05.003}}.

\bibitem{grohol2024predictive}
C.~M. Grohol, Y.~C. Shin, {On the predictive accuracy and sensitivity to
  property variations in the high-fidelity modeling of the molten pool dynamics
  in laser melting of metals}, International Journal of Heat and Mass Transfer
  232~(June) (2024) 125894.
\newblock \href {https://doi.org/10.1016/j.ijheatmasstransfer.2024.125894}
  {\path{doi:10.1016/j.ijheatmasstransfer.2024.125894}}.

\bibitem{yu2024mechanism}
T.~Yu, S.~Zhao, J.~Zhao, {A mechanism-based optimization strategy with adaptive
  laser power for laser powder bed fusion}, Additive Manufacturing 92~(August)
  (2024) 104403.
\newblock \href {https://doi.org/10.1016/j.addma.2024.104403}
  {\path{doi:10.1016/j.addma.2024.104403}}.

\bibitem{geiger20093d}
M.~Geiger, K.-H. Leitz, H.~Koch, A.~Otto, {A 3D transient model of keyhole and
  melt pool dynamics in laser beam welding applied to the joining of zinc
  coated sheets}, Production Engineering 3~(2) (2009) 127--136.
\newblock \href {https://doi.org/10.1007/s11740-008-0148-7}
  {\path{doi:10.1007/s11740-008-0148-7}}.

\bibitem{chen2020spattering}
H.~Chen, W.~Yan, {Spattering and denudation in laser powder bed fusion process:
  Multiphase flow modelling}, Acta Materialia 196 (2020) 154--167.
\newblock \href {https://doi.org/10.1016/j.actamat.2020.06.033}
  {\path{doi:10.1016/j.actamat.2020.06.033}}.

\bibitem{lee2015mesoscopic}
Y.~S. Lee, W.~Zhang, {Mesoscopic simulation of heat transfer and fluid flow in
  laser powder bed additive manufacturing}, in: Solid Free Form Fabrication
  Symposium, University of Texas at Austin, Austin, Texas, USA, 2015, pp.
  1154--1165.

\bibitem{zakirov2024kissam}
A.~Zakirov, S.~Belousov, M.~Bogdanova, B.~Korneev, I.~Iskandarova,
  A.~Perepelkina, B.~Potapkin, {KiSSAM: efficient simulation of melt pool
  dynamics during PBF using GPUs}, Progress in Additive Manufacturing 9~(5)
  (2024) 1491--1508.
\newblock \href {https://doi.org/10.1007/s40964-023-00561-1}
  {\path{doi:10.1007/s40964-023-00561-1}}.

\bibitem{ikeda2025high}
K.~Ikeda, S.~Sakane, T.~Aoki, T.~Takaki, {High-performance phase-field lattice
  Boltzmann simulations for accurate thermal fluid flow in metal additive
  manufacturing}, IOP Conference Series: Materials Science and Engineering
  1335~(1) (2025) 012016.
\newblock \href {https://doi.org/10.1088/1757-899X/1335/1/012016}
  {\path{doi:10.1088/1757-899X/1335/1/012016}}.

\bibitem{ammer2014simulating}
R.~Ammer, M.~Markl, U.~Ljungblad, C.~K{\"{o}}rner, U.~R{\"{u}}de, {Simulating
  fast electron beam melting with a parallel thermal free surface lattice
  Boltzmann method}, Computers \& Mathematics with Applications 67~(2) (2014)
  318--330.
\newblock \href {https://doi.org/10.1016/j.camwa.2013.10.001}
  {\path{doi:10.1016/j.camwa.2013.10.001}}.

\bibitem{meier2021novel}
C.~Meier, S.~L. Fuchs, A.~J. Hart, W.~A. Wall, {A novel smoothed particle
  hydrodynamics formulation for thermo-capillary phase change problems with
  focus on metal additive manufacturing melt pool modeling}, Computer Methods
  in Applied Mechanics and Engineering 381 (2021) 113812.
\newblock \href {https://doi.org/10.1016/j.cma.2021.113812}
  {\path{doi:10.1016/j.cma.2021.113812}}.

\bibitem{fuchs2022versatile}
S.~L. Fuchs, P.~M. Praegla, C.~J. Cyron, W.~A. Wall, C.~Meier, {A versatile SPH
  modeling framework for coupled microfluid-powder dynamics in additive
  manufacturing: binder jetting, material jetting, directed energy deposition
  and powder bed fusion}, Engineering with Computers 38~(6) (2022) 4853--4877.
\newblock \href {https://doi.org/10.1007/s00366-022-01724-4}
  {\path{doi:10.1007/s00366-022-01724-4}}.

\bibitem{luthi2023adaptive}
C.~L{\"{u}}thi, M.~Afrasiabi, M.~Bambach, {An adaptive smoothed particle
  hydrodynamics (SPH) scheme for efficient melt pool simulations in additive
  manufacturing}, Computers \& Mathematics with Applications 139 (2023) 7--27.
\newblock \href {https://doi.org/10.1016/j.camwa.2023.03.003}
  {\path{doi:10.1016/j.camwa.2023.03.003}}.

\bibitem{lin2023enhanced}
Y.~Lin, C.~L{\"{u}}thi, M.~Afrasiabi, M.~Bambach, {Enhanced heat source
  modeling in particle-based laser manufacturing simulations with ray tracing},
  International Journal of Heat and Mass Transfer 214 (2023) 124378.
\newblock \href {https://doi.org/10.1016/j.ijheatmasstransfer.2023.124378}
  {\path{doi:10.1016/j.ijheatmasstransfer.2023.124378}}.

\bibitem{ma2025gpu}
Y.~Ma, Z.~Zhang, C.~Wei{\ss}enfels, M.~Zhou, L.~Ma, X.~Tang, M.~Liu,
  {GPU-accelerated multi-phase, multi-resolution SPH method with ray tracing
  for laser powder bed fusion}, Computer Methods in Applied Mechanics and
  Engineering 448~(July 2025) (2025) 118423.
\newblock \href {https://doi.org/10.1016/j.cma.2025.118423}
  {\path{doi:10.1016/j.cma.2025.118423}}.

\bibitem{russell2018numerical}
M.~Russell, A.~Souto-Iglesias, T.~Zohdi, {Numerical simulation of Laser Fusion
  Additive Manufacturing processes using the SPH method}, Computer Methods in
  Applied Mechanics and Engineering 341 (2018) 163--187.
\newblock \href {https://doi.org/10.1016/j.cma.2018.06.033}
  {\path{doi:10.1016/j.cma.2018.06.033}}.

\bibitem{anisimov1995instabilities}
S.~Anisimov, V.~Khokhlov, {Instabilities in Laser-Matter Interaction}, CRC
  Press, Boca Raton, FL, 1995.

\bibitem{ross2022volumetric}
A.~J. Ross, I.~Bitharas, K.~G. Perkins, A.~J. Moore, {Volumetric heat source
  calibration for laser powder bed fusion}, Additive Manufacturing 60 (2022).
\newblock \href {https://doi.org/10.1016/J.ADDMA.2022.103267}
  {\path{doi:10.1016/J.ADDMA.2022.103267}}.

\bibitem{olsson2007conservative}
E.~Olsson, G.~Kreiss, S.~Zahedi, {A conservative level set method for two phase
  flow II}, Journal of Computational Physics 225~(1) (2007) 785--807.
\newblock \href {https://doi.org/10.1016/j.jcp.2006.12.027}
  {\path{doi:10.1016/j.jcp.2006.12.027}}.

\bibitem{nitsche1971uber}
J.~Nitsche, {{\"{U}}ber ein Variationsprinzip zur L{\"{o}}sung von
  Dirichlet-Problemen bei Verwendung von Teilr{\"{a}}umen, die keinen
  Randbedingungen unterworfen sind}, Abhandlungen aus dem Mathematischen
  Seminar der Universit{\"{a}}t Hamburg 36~(1) (1971) 9--15.
\newblock \href {https://doi.org/10.1007/BF02995904}
  {\path{doi:10.1007/BF02995904}}.

\bibitem{kronbichler2012generic}
M.~Kronbichler, K.~Kormann, {A generic interface for parallel cell-based finite
  element operator application}, Computers \& Fluids 63 (2012) 135--147.
\newblock \href {https://doi.org/10.1016/j.compfluid.2012.04.012}
  {\path{doi:10.1016/j.compfluid.2012.04.012}}.

\bibitem{arndt2025deal}
D.~Arndt, W.~Bangerth, M.~Bergbauer, B.~Blais, M.~Fehling, R.~Gassm{\"{o}}ller,
  T.~Heister, L.~Heltai, M.~Kronbichler, M.~Maier, P.~Munch, S.~Scheuerman,
  B.~Turcksin, S.~Uzunbajakau, D.~Wells, M.~Wichrowski, {The deal.II library,
  version 9.7}, Journal of Numerical Mathematics 32~(4) (2025) 369--380.
\newblock \href {https://doi.org/10.1515/jnma-2025-0115}
  {\path{doi:10.1515/jnma-2025-0115}}.

\bibitem{kronbichler2018fast}
M.~Kronbichler, A.~Diagne, H.~Holmgren, {A fast massively parallel two-phase
  flow solver for microfluidic chip simulation}, The International Journal of
  High Performance Computing Applications 32~(2) (2018) 266--287.
\newblock \href {https://doi.org/10.1177/1094342016671790}
  {\path{doi:10.1177/1094342016671790}}.

\bibitem{sussman1994level}
M.~Sussman, P.~Smereka, S.~Osher, {A Level Set Approach for Computing Solutions
  to Incompressible Two-Phase Flow}, Journal of Computational Physics 114~(1)
  (1994) 146--159.
\newblock \href {https://doi.org/10.1006/jcph.1994.1155}
  {\path{doi:10.1006/jcph.1994.1155}}.

\bibitem{peskin2002immersed}
C.~S. Peskin, {The immersed boundary method}, Acta Numerica 11 (2002) 479--517.
\newblock \href {https://doi.org/10.1017/S0962492902000077}
  {\path{doi:10.1017/S0962492902000077}}.

\bibitem{knight1979theoretical}
C.~J. Knight, {Theoretical Modeling of Rapid Surface Vaporization with Back
  Pressure}, AIAA Journal 17~(5) (1979) 519--523.
\newblock \href {https://doi.org/10.2514/3.61164} {\path{doi:10.2514/3.61164}}.

\bibitem{voller1987fixed}
V.~Voller, C.~Prakash, {A fixed grid numerical modelling methodology for
  convection-diffusion mushy region phase-change problems}, International
  Journal of Heat and Mass Transfer 30~(8) (1987) 1709--1719.
\newblock \href {https://doi.org/10.1016/0017-9310(87)90317-6}
  {\path{doi:10.1016/0017-9310(87)90317-6}}.

\bibitem{schreter2024consistent}
M.~Schreter-Fleischhacker, P.~Munch, N.~Much, M.~Kronbichler, W.~A. Wall,
  C.~Meier, {A consistent diffuse-interface model for two-phase flow problems
  with rapid evaporation}, Advanced Modeling and Simulation in Engineering
  Sciences 11~(1) (2024) 19.
\newblock \href {https://doi.org/10.1186/s40323-024-00276-0}
  {\path{doi:10.1186/s40323-024-00276-0}}.

\bibitem{ludvigsson2018high}
G.~Ludvigsson, K.~R. Steffen, S.~Sticko, S.~Wang, Q.~Xia, Y.~Epshteyn,
  G.~Kreiss, {High-Order Numerical Methods for 2D Parabolic Problems in Single
  and Composite Domains}, Journal of Scientific Computing 76~(2) (2018)
  812--847.
\newblock \href {https://doi.org/10.1007/s10915-017-0637-y}
  {\path{doi:10.1007/s10915-017-0637-y}}.

\bibitem{burman2015cutfem}
E.~Burman, S.~Claus, P.~Hansbo, M.~G. Larson, A.~Massing, {CutFEM: Discretizing
  geometry and partial differential equations}, International Journal for
  Numerical Methods in Engineering 104~(7) (2015) 472--501.
\newblock \href {https://doi.org/10.1002/nme.4823}
  {\path{doi:10.1002/nme.4823}}.

\bibitem{benzaken2024contructing}
J.~Benzaken, J.~A. Evans, R.~Tamstorf, {Constructing Nitsche's Method for
  Variational Problems}, Archives of Computational Methods in Engineering
  31~(4) (2024) 1867--1896.
\newblock \href {https://doi.org/10.1007/s11831-023-09953-6}
  {\path{doi:10.1007/s11831-023-09953-6}}.

\bibitem{sticko2016stabilized}
S.~Sticko, G.~Kreiss, {A stabilized Nitsche cut element method for the wave
  equation}, Comput. Methods Appl. Mech. Engrg 309 (2016) 364--387.
\newblock \href {https://doi.org/10.1016/j.cma.2016.06.001}
  {\path{doi:10.1016/j.cma.2016.06.001}}.

\bibitem{burman2012fictitious}
E.~Burman, P.~Hansbo, {Fictitious domain finite element methods using cut
  elements: II. A stabilized Nitsche method}, Applied Numerical Mathematics
  62~(4) (2012) 328--341.
\newblock \href {https://doi.org/10.1016/j.apnum.2011.01.008}
  {\path{doi:10.1016/j.apnum.2011.01.008}}.

\bibitem{lee2003immersed}
L.~Lee, R.~J. LeVeque, {An Immersed Interface Method for Incompressible
  Navier--Stokes Equations}, SIAM Journal on Scientific Computing 25~(3) (2003)
  832--856.
\newblock \href {https://doi.org/10.1137/S1064827502414060}
  {\path{doi:10.1137/S1064827502414060}}.

\bibitem{schreter2023evaluation}
M.~Schreter-Fleischhacker, P.~Munch,
  \href{https://dealii.org/developer/doxygen/deal.II/step_87.html}{{The deal.II
  tutorial step-87: Evaluation of finite element solutions at arbitrary points
  within a distributed mesh with application to two-phase flow}} (2023).
\newline\urlprefix\url{https://dealii.org/developer/doxygen/deal.II/step_87.html}

\bibitem{lorensen1987marching}
W.~E. Lorensen, H.~E. Cline, {Marching cubes: A high resolution 3D surface
  construction algorithm}, in: Proceedings of the 14th annual conference on
  Computer graphics and interactive techniques, Vol.~21, ACM, New York, NY,
  USA, 1987, pp. 163--169.
\newblock \href {https://doi.org/10.1145/37401.37422}
  {\path{doi:10.1145/37401.37422}}.

\bibitem{kothe1996volume}
D.~Kothe, W.~Rider, S.~Mosso, J.~Brock, J.~Hochstein, {Volume tracking of
  interfaces having surface tension in two and three dimensions}, in: 34th
  Aerospace Sciences Meeting and Exhibit, American Institute of Aeronautics and
  Astronautics, Reston, Virigina, 1996, p. 859.
\newblock \href {https://doi.org/10.2514/6.1996-859}
  {\path{doi:10.2514/6.1996-859}}.

\bibitem{yokoi2014density}
K.~Yokoi, {A density-scaled continuum surface force model within a balanced
  force formulation}, Journal of Computational Physics 278~(1) (2014) 221--228.
\newblock \href {https://doi.org/10.1016/j.jcp.2014.08.034}
  {\path{doi:10.1016/j.jcp.2014.08.034}}.

\bibitem{zahedi2012spurious}
S.~Zahedi, M.~Kronbichler, G.~Kreiss, {Spurious currents in finite element
  based level set methods for two‐phase flow}, International Journal for
  Numerical Methods in Fluids 69~(9) (2012) 1433--1456.
\newblock \href {https://doi.org/10.1002/fld.2643}
  {\path{doi:10.1002/fld.2643}}.

\bibitem{francois2006balanced}
M.~M. Francois, S.~J. Cummins, E.~D. Dendy, D.~B. Kothe, J.~M. Sicilian, M.~W.
  Williams, {A balanced-force algorithm for continuous and sharp interfacial
  surface tension models within a volume tracking framework}, Journal of
  Computational Physics 213~(1) (2006) 141--173.
\newblock \href {https://doi.org/10.1016/j.jcp.2005.08.004}
  {\path{doi:10.1016/j.jcp.2005.08.004}}.

\bibitem{schott2017stabilized}
B.~Schott, {Stabilized Cut Finite Element Methods for Complex Interface Coupled
  Flow Problems}, Ph.D. thesis, Technical University of Munich (2017).

\bibitem{saye2015high}
R.~I. Saye, {High-Order Quadrature Methods for Implicitly Defined Surfaces and
  Volumes in Hyperrectangles}, SIAM Journal on Scientific Computing 37~(2)
  (2015) A993--A1019.
\newblock \href {https://doi.org/10.1137/140966290}
  {\path{doi:10.1137/140966290}}.

\bibitem{bergbauer2025high}
M.~Bergbauer, P.~Munch, W.~A. Wall, M.~Kronbichler, {High-Performance
  Matrix-Free Unfitted Finite Element Operator Evaluation}, SIAM Journal on
  Scientific Computing 47~(3) (2025) B665--B689.
\newblock \href {https://doi.org/10.1137/24M1653689}
  {\path{doi:10.1137/24M1653689}}.

\bibitem{boivineau2006thermophysical}
M.~Boivineau, C.~Cagran, D.~Doytier, V.~Eyraud, M.~H. Nadal, B.~Wilthan,
  G.~Pottlacher, {Thermophysical Properties of Solid and Liquid Ti-6Al-4V
  (TA6V) Alloy}, International Journal of Thermophysics 27~(2) (2006) 507--529.
\newblock \href {https://doi.org/10.1007/PL00021868}
  {\path{doi:10.1007/PL00021868}}.

\bibitem{mohr2020precise}
M.~Mohr, R.~Wunderlich, R.~Novakovic, E.~Ricci, H.-J. Fecht, {Precise
  Measurements of Thermophysical Properties of Liquid Ti–6Al–4V (Ti64)
  Alloy On Board the International Space Station}, Advanced Engineering
  Materials 22~(7) (2020).
\newblock \href {https://doi.org/10.1002/adem.202000169}
  {\path{doi:10.1002/adem.202000169}}.

\bibitem{zhang2020element}
G.~Zhang, J.~Chen, M.~Zheng, Z.~Yan, X.~Lu, X.~Lin, W.~Huang, {Element
  Vaporization of Ti-6Al-4V Alloy during Selective Laser Melting}, Metals
  10~(4) (2020) 435.
\newblock \href {https://doi.org/10.3390/met10040435}
  {\path{doi:10.3390/met10040435}}.

\bibitem{ma2011direct}
C.~Ma, D.~Bothe, {Direct numerical simulation of thermocapillary flow based on
  the Volume of Fluid method}, International Journal of Multiphase Flow 37~(9)
  (2011) 1045--1058.
\newblock \href {https://doi.org/10.1016/j.ijmultiphaseflow.2011.06.005}
  {\path{doi:10.1016/j.ijmultiphaseflow.2011.06.005}}.

\bibitem{minissale2018effect}
M.~Minissale, G.~T. Zeweldi, R.~Bisson, L.~Gallais, {The effect of surface
  temperature on optical properties of molybdenum mirrors in the visible and
  near-infrared domains}, Nuclear Fusion 58~(9) (2018) 096012.
\newblock \href {https://doi.org/10.1088/1741-4326/aaca03}
  {\path{doi:10.1088/1741-4326/aaca03}}.

\bibitem{simonds2021causal}
B.~J. Simonds, J.~Tanner, A.~Artusio-Glimpse, P.~A. Williams, N.~Parab,
  C.~Zhao, T.~Sun, {The causal relationship between melt pool geometry and
  energy absorption measured in real time during laser-based manufacturing},
  Applied Materials Today 23 (2021) 101049.
\newblock \href {https://doi.org/10.1016/j.apmt.2021.101049}
  {\path{doi:10.1016/j.apmt.2021.101049}}.

\end{thebibliography}

\end{document}